\documentclass[11pt,preprint]{aastex}

\shorttitle{Line strengths of early-type galaxies.}
\shortauthors{OGANDO ET AL.}

\begin{document}


\title{Line strengths of early-type galaxies.\altaffilmark{1}}
\author{Ricardo L.C. Ogando,\altaffilmark{2,3} Marcio A.G. Maia,\altaffilmark{2}
Paulo S. Pellegrini,\altaffilmark{2} and Luiz N. da
Costa\altaffilmark{2}}

\altaffiltext{1}{Partly based on observations at European Southern
Observatory (ESO) at the 1.52m telescope under the ESO-ON
agreement.}

\altaffiltext{2}{Observat\'orio Nacional (ON/MCT), Rua General
Jos\'e Cristino,~77,  Rio de Janeiro 20921-400 - RJ, Brazil;
maia@on.br, pssp@on.br,ldacosta@on.br}

\altaffiltext{3}{Instituto de F\'\i sica, Universidade Federal do
Rio de Janeiro, (RJ), Brazil; ogando@if.ufrj.br}


\begin{abstract}
In this paper we present measurements of velocity dispersions and Lick indices for 509 galaxies in the local Universe, based on high signal-to-noise, long slit spectra obtained with the 1.52 m ESO telescope at La Silla. The conversion of our measurements into the Lick/IDS system was carried out following the general prescription of \citet{Wor97}. Comparisons of our measurements with those of other authors show, in general, good agreement. We also examine the dependence between these indices (e.g., H$\beta$, Mg$_2$, Fe5270 and NaD) and the central velocity dispersion ($\sigma_0$), and we find that they are consistent with those previously reported in the literature. Benefiting from the relatively large size of the sample, we are able to investigate the dependence of these relations on morphology and environment, here represented by the local galaxy density. We find that for metallic lines these relations show no significant dependence on environment or morphology, except in the case of NaD, which shows distinct behavior for E and S0. On the other hand, the H$\beta$-$\log\sigma_0$ shows a significant difference as a function of the local density of galaxies, which we interpret as being caused by the truncation of star formation in high density environments. Comparing our results with those obtained by other authors we find a few discrepancies, adding to the ongoing debate about the nature of these relations. Finally, we report that the scatter of the Mg indices versus $\sigma_0$ relations correlate with H$\beta$, suggesting that age may contribute to the scatter. Furthermore, this scatter shows no significant dependence on morphology or environment. Our results are consistent with the currently downsizing model, where low mass galaxies have an extended star formation history, except for those located in high density regions.

\end{abstract}

\keywords{galaxies: formation --- galaxies: elliptical and lenticular --- galaxies: stellar content}

\section{Introduction}
\label{intro}

Efforts to understand how early-type galaxies formed and evolved
have shown the importance of determining a group of fundamental
parameters and investigate their relations. For instance, the study of their
central velocity dispersions ($\sigma_0$), luminosities ($L$) and
characteristic sizes like the effective radius ($r_e$), which
reflect the influence of both dark and baryonic matter, provided
in the last decades a deeper understanding of the structural and
kinematic properties of this class of galaxies. Moreover,
relations among these parameters such as those referred as the
Faber-Jackson \citep{Fab76} and the Fundamental Plane
\citep{DjDv87, Dre87} proved to be important tools to study galaxy formation and evolution. In addition, their chemical enrichment can be addressed by
means of colors or absorption line features of key elements such
as Mg, Fe and H Balmer lines. These can be compared to predictions
of theoretical evolutionary single stellar population (SSP) models
\citep[e.g.,][]{Wor94,Tho03}, which summarize galaxies stellar
content properties, such as age and metallicity, along with
distinct elements abundance ratios. In practice, line strengths
have some advantages over colors because they are less affected by
dust effects and are useful in breaking the so called
age-metallicity degeneracy, which prevents us from delivering a
straight interpretation of galaxy light.

A particularly good example of the insight provided by line
strengths analysis, as well as the difficulties involved, can be
seen from the analysis of the well known Mg$_2-\sigma_0$ relation
\citep[e.g.,][]{Gor90,Ben93,Ber98}. This is usually interpreted as
a mass-metalicity relation, since more massive galaxies are, in
principle, able to keep larger amounts of enriched gas blown out
in supernovae events, thus increasing the metallicity of the
material from which new stellar generations are formed. 
By the same token, this class of objects should have an extended history
of star formation, as predicted by both traditional scenarios of galaxy formation: the monolithic dissipative collapse \citep{Lar74} and the hierarchical clustering \citep{Kau93}, implying that the mass-metalicity relation should also be subject to age. 
Indeed, while this is, to some extent, true for early-type galaxies in general
\citep{Tra00}, in the case of massive objects, observational
evidences conflict with predictions, because they are essentially
old \citep{Kun00} and had a short star formation time scale, as
indicated by the super-solar $\alpha$/Fe ratio \citep{Wor92,Tho05}. 
On the other hand, low mass objects have an
extended star formation history, which is currently known as
downsizing \citep{Cow96}, or yet anti-hierarchical, since in the
hierarchical clustering scenario they should be the first galaxies
to appear and therefore be the oldest. Part of the answer to that
inconsistency might be related to the details of the baryonic
matter interactions, in the form of dissipation and galactic
winds, driven by supernovae or active galactic nuclei
\citep[e.g.,][]{Pip04,GdL06}.

While the Mg$_2-\sigma_0$ relation has proven to be extremely useful, the measurement of other line indices can provide additional information such as age, formation time-scale, and abundance, that may help discriminate among competing models of galaxy formation.
The main goal of the present paper is to provide measurements for 10 Lick indices for a representative sample of nearby early-type galaxies 
in order to characterize their dominant stellar population.
For that purpose we use photometric and spectroscopic data for a subset of nearby early-type galaxies drawn from the ENEAR sample \citep{daC00} – as well as new higher S/N spectroscopic data.
These data are used to investigate the dependence of the index$-\log\sigma_0$ on morphology (E and S0) and environment.

The present sample complements those recently available at intermediate redshifts such as the NOAO Fundamental Plane Survey \citep[e.g.,][]{Nel05} and the SDSS, now a standard within z=0.3 \citep[e.g.,][]{Ber05}. 
While existing low redshift samples \citep[e.g.,][]{Tho05,San06a} are relatively small compared to those at intermediate redshifts, they have the advantage of having higher quality spectroscopic data and more reliable  morphological and structural information 

This paper is organized as follows: in \S\ \ref{sample} we describe the sample selection and the observations and in \S\ \ref{redu}, the data reduction. 
In \S\ \ref{mea_vel_disp} we present the velocity dispersion measurements.
The methodology used to compute line indices and the results of their comparison with those available in the literature are presented in \S\ \ref{line_meas}. In \S\ \ref{index_sigma} we analyze the index$-\log\sigma_0$ relations
and their dependence on galaxy morphology and environment.
Finally, in \S\ \ref{disc_conc} we present a summary of our conclusions.

%
\section{The Sample and Observations}
\label{sample}
The sample discussed here consists of objects extracted from the
ENEAR catalog \citep{daC00}. This survey contains a database of
photometric \citep{Alo03} and spectroscopic \citep{Weg03}
parameters for a magnitude limited ($m_B\leq14.5$) sample of
early-type galaxies which is considered representative of the
nearby universe ($v_r< 7000$\ km s$^{-1}$). Several global
parameters are available in this database such as magnitudes
($m_R$), the effective radius, mean surface brightness ($\mu_e$) within $r_{e}$, characteristic diameter ($D_n$), and the central velocity dispersion.

\begin{figure}
   \includegraphics[angle=-90,scale=0.5]{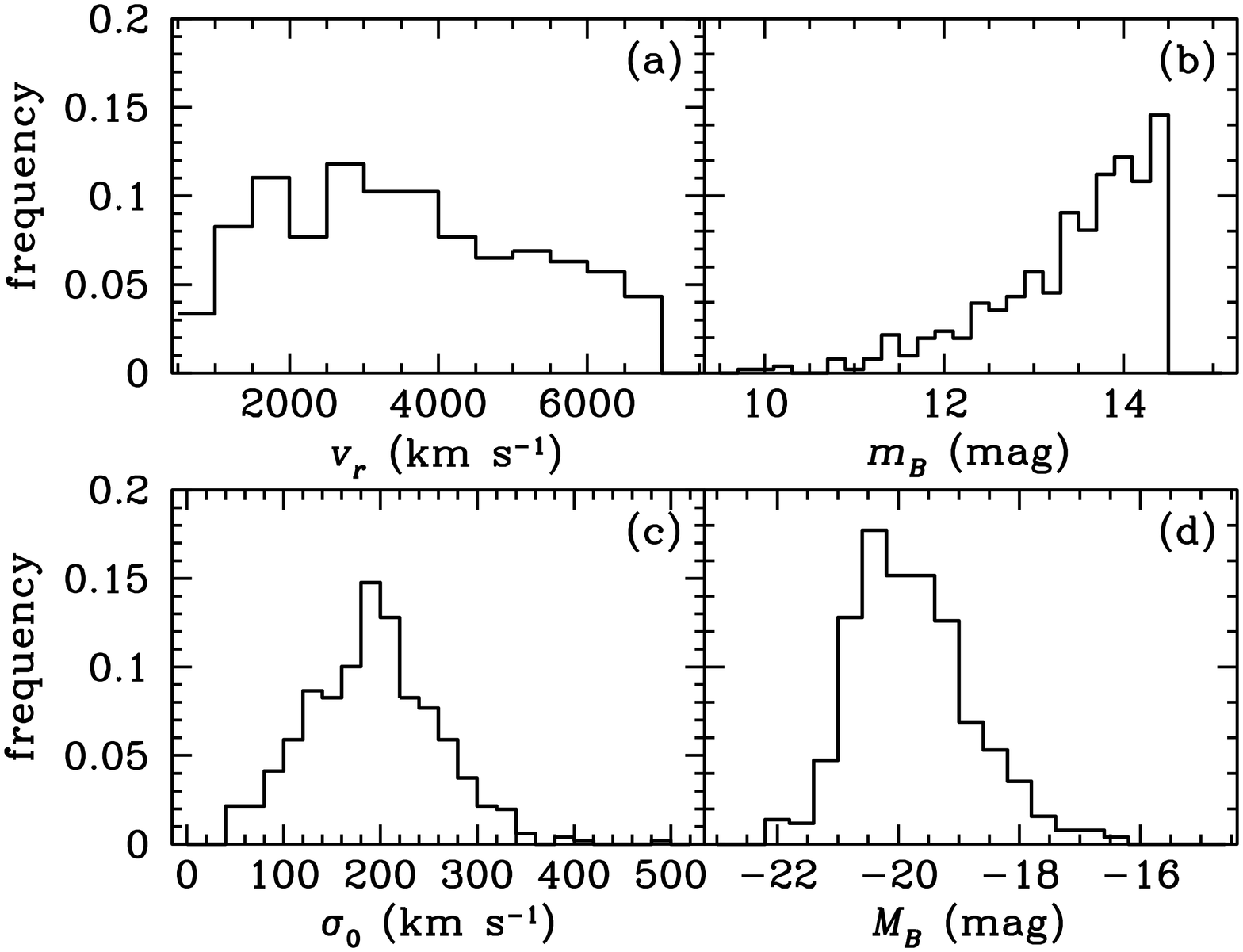}
\vspace{0.0 cm}
   \caption{Distributions of $v_r$, $m_B$, $\sigma_0$ and $M_B$ for the galaxies of our sample.
   }
   \label{vel_mag_Mag_sig_dist}
\end{figure}

This sample includes objects residing in different environments such as the general field, groups and clusters.
As an additional criteria we excluded objects with $\mid
b\mid\lesssim 15^\circ$ to avoid the galactic plane, 
and also objects with $\delta\gtrsim 30^\circ$ 
due to the observational constraints. The observed list consists of 115 E, 131 E/S0
and 263 S0, giving a total of 509 galaxies, whose physical properties (L,
$r_{e}$, $\sigma_0$) are fairly representative of the parent
sample. In this paper we group, for analysis purposes, the
galaxies E and E/S0 in the general class of E galaxies. These
morphological types were revised based on visual inspection in two
bands using images available in the Digital Sky Survey
\citep{Las90}. In Table~\ref{tab:lisgal} we display the catalog of
observed galaxies, including some information about their physical
properties. The columns in this catalog refer to: (1) object name;
(2) and (3) are for (J2000.0) equatorial coordinates; (4)
morphological type $T$; (5) apparent $B$-band magnitude $m_B$; (6)
the heliocentric radial velocity $v_r$ and respective error in km
s$^{-1}$; (7) the logarithm of $\sigma_0$ and its associated error
in km s$^{-1}$ from our determinations (see \S\ 
\ref{mea_vel_disp}); (8) the logarithm of $r_e$ in units of kpc,
based on the data of \cite{Alo03} and the D$_n-\sigma$ distances
given by \citet{Ber02}. Finally, in column (9), the number of companions $N_c$ as defined in \S\ \ref{gal_env} is presented. To give an idea of the global sample characteristics, the $v_r$, $m_B$, $M_B$ and $\sigma_0$ distributions are displayed in panels of Figure \ref{vel_mag_Mag_sig_dist}.

The line strengths presented in this paper are based on long-slit spectra collected using the Boller \& Chivens spectrograph mounted on the 1.52 m telescope at the La Silla site of the European Southern Observatory, during several runs between November 1993 and November 2002. 
Our observations were carried out with a 4.1$''\times 2.5''$ slit centered on the galaxy nucleus using a TV guider.
For the vast majority of the objects the spectrograph was rotated to allow the slit to be oriented along the galaxy's major axis. For some galaxies, observations were also done along the minor axis, and for a few of them the slit was oriented in an intermediate position. 
Typical exposure time was 10 minutes in the beginning of the survey, and 30 minutes at a later phase. More than one frame per object was frequently obtained, in particular, for a set of $\sim 140$ bright galaxies of the ENEAR sample selected to study metallicity gradients.

\begin{figure}
   \includegraphics[angle=0,scale=0.5]{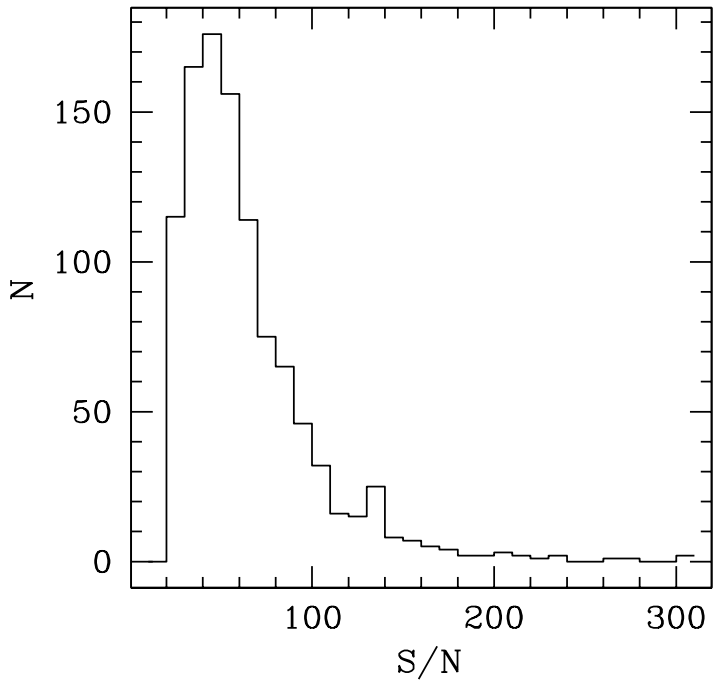}
\vspace{0.0 cm}
   \caption{Distribution of $S/N$ for the galaxies of our sample.}
   \label{dist_sn_global}
\end{figure}

The various instrumental setups used are described in
Table~\ref{tab:setups}. The spectra were considered acceptable for
the present analysis only if they had $S/N > 20$. The $S/N$ was determined in the continuum range from 5860 \AA\ to 5875 \AA\ as the ratio between the average and the $rms$ of the signal in that region. 
In Figure \ref{dist_sn_global} we display the distribution of $S/N$ for the observed galaxies. The typical resolutions are 3 \AA\ and 6 \AA\ (FWHM) as discussed below.

Dome flats and bias were taken on a nightly basis. The dark
current was found to be negligible after all, and its correction
was not applied. After each science target exposure, at the same
position of the sky that the object was observed, arc lamps (He-Ar
or He-Ar-Fe-Ne) were exposed for wavelength calibration.
In addition, we observed radial velocity standard stars and stars
in common with Lick/IDS system, which are mainly in the F to K spectral
type range, with luminosity classes ranging from I to V. To
determine the velocity dispersion of galaxies we restricted the
sample of stars to G and K giants. Besides the calibration to the
Lick/IDS system, all those standards allowed calibrations among
different setups.

\section{Data Reduction}
\label{redu}

We followed standard procedures to reduce CCD spectra using
\texttt{IRAF}. The science frames were subtracted by an average
bias frame and divided by a normalized average flat-field frame.
The 2-D images had cosmic rays hits in the surroundings of the
spectrum removed with the task \texttt{imedit}. Those hits 
reaching the frame in the region next to the spectrum were not 
removed to avoid the introduction of artificial features. We also 
decided to discard the equivalent widths of lines which had cosmic 
rays within their definition range.

The extraction of 1-D spectrum was made with the task
\texttt{apall} setting the aperture to obtain most of the galaxy's
light along the slit as possible. This means that the aperture
size typically reached a radial distance from the galaxy's center
where the intensity is $\approx$ 5\% of the peak value of the
light profile. With this procedure we intended to obtain spectra
containing as much information as possible of the galaxy as a
whole, and not just from its central region.

Arc lamps observed after each object frame were used to provide
wavelength solutions and spectra linearization. We adopted a
5-7$^{th}$ order Legendre polynomial and the final fits used
$\sim$ 20 and 50 lines with typical $rms$ $\sim$ 0.02 and 0.08
\AA, for the 1200 l mm$^{-1}$ and 600 l mm$^{-1}$ gratings,
respectively. The wavelength solution was verified by checking that the sky lines (e.g., [O {\sc{i}}]$\lambda$5577\ \AA) were located at their expected rest wavelength. 
The spectral resolution is $\sim 3$ \AA\ or $\sim 6$ \AA\ depending on
the grating used (see Table~\ref{tab:setups}). No flux calibration
was applied to the spectra.

\section{Measurement of $v_r$ and $\sigma_0$}
\label{mea_vel_disp}

Radial velocity and velocity dispersions were measured for each
spectrum using the cross-correlation technique \citep{Ton79}. In
particular, we used the \texttt{RVSAO} package \citep{Kur98} to
obtain $v_r$ by cross-correlating galaxy spectra with several
templates. The velocity and error adopted was provided by the best
template (largest correlation coefficient). Typical error for the
$v_r$ measurement is $\sim 20$ km s$^{-1}$.

For the measurement of $\sigma_0$ we generated a series of
templates for each run/grating, combining stellar spectra of G and
K giants since they represent the absorption features 
present in early-type galaxy spectra quite well. We used the wavelength 
interval from 4500 \AA\ to 6000 \AA, which is common to all the 
setups, to determine $\sigma_0$. Although this range encompasses
the H$\beta$ and NaD lines, which are suspected of providing poor 
fits, we tested different ranges, excluding one or both of them, 
finding no significant difference.

Another issue addressed in our procedure is the fact that the velocity dispersion varies with galactocentric radius.
Global apertures, like the ones we are using,
do not sample systematically a characteristic size of the galaxy
like e.g., some fraction of $r_e$, or some fixed size at the
galaxy's referential frame. To avoid this bias, we
followed the usual $\sigma_0$ aperture correction method,
considering the galaxy at some fiducial distance (Coma cluster)
and using a standard metric aperture. Following the prescription
of \citet{Jor95}, the aperture correction is given by the equation:
\begin{equation}
\log(\sigma_{norm})=\log(\sigma_{ab}) - \alpha
\log(r_{ab}/r_{norm}) \label{sigma_cor_Jor}
\end{equation}
\noindent where $\alpha$ is the average gradient of $\log\sigma$,
$r_{ab}$ and $r_{norm}$ are the sizes of circular and normalized
apertures, respectively. The $r_{ab}$ radius is given by
$1.025[lc/\pi]^{1/2}$, where $l$ is the slit width and, $c$ is the
effective length of the rectangular aperture to simulate a
circular one. We choose a metric aperture of 1.19 kpc
\citep{Dav87} and $\alpha=-0.04$ \citep{Jor95}.

For galaxies with multiple observations, we calculated an average
value for $\sigma_0$ in two steps, first including all
measurements, then a final value is computed discarding individual
$\sigma_0$ greater than two times the $rms$ limits from the
average. The typical $\sigma_0$ error is $\sim17$ km s$^{-1}$.
Also, for the low resolution ($\sim 6$ \AA) configurations, we had
to avoid $\sigma_0$ determinations below 160 km s$^{-1}$, as given
by their high resolution counterpart, because low resolution
measurements of low $\sigma_0$ galaxies tend to overestimate it.

\begin{figure}
   \includegraphics[angle=0,scale=0.4]{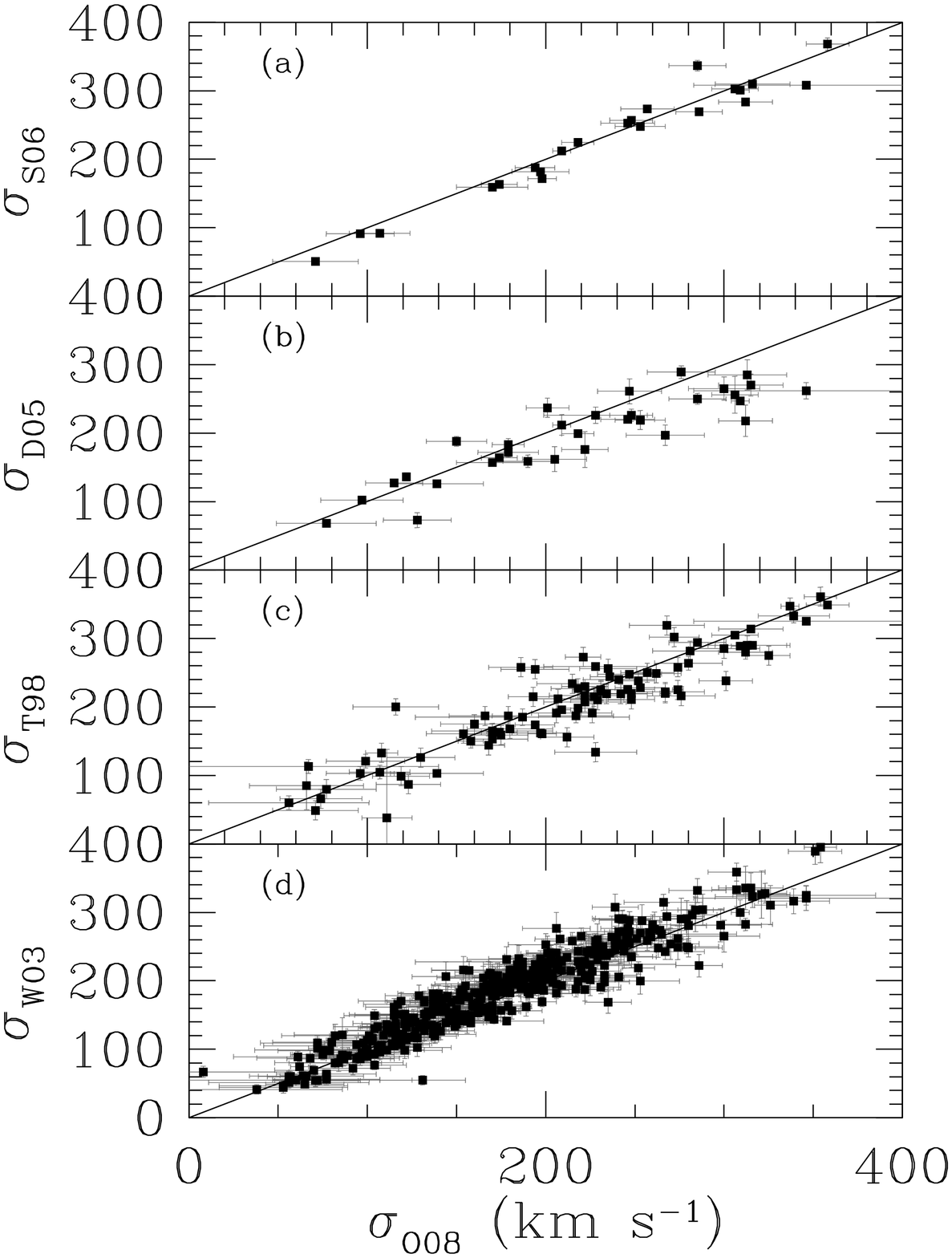}
   \vspace{0.0 cm}
   \caption{Comparison between our measurements of $\sigma_0$ ($\sigma_{O08}$) and those
   values by \citet{San06a} (panel a); \citet{Den05a} (panel b); \citet{Tra98} (panel c) and
   \citet{Weg03} (panel d). }
   \label{sigcompfig}
\end{figure}

To check the quality of our $\sigma_0$ measurements, we
performed a comparison with galaxies in common with \citet{Tra98},
\citet{Den05a} and \citet{San06a}.
We also compare these new 
measurements with those in the original ENEAR paper presented by \citet{Weg03}. 
The results are presented in Figure \ref{sigcompfig} and Table \ref{tab:vdautcmp},
where we show the number of galaxies in common with each author, the mean offset 
$\langle\delta\sigma_0\rangle$, its $rms$, and the $t$ value given by the Student $t$ test, where $t>1.96$ means that the offset is significant at 95\% confidence level.
In general, we have a good agreement with \citet{San06a} who observed
with instrumental resolutions similar to ours. The small
difference we find with \citet{Den05a} is probably caused by
their lower instrumental resolution of 6 \AA. 
\citet{Tra98} measurements also shows good agreement over the entire
range of $\sigma_0$. Most of the data they present, came from
\citet{Dav87}, who used an instrumental resolution up to 85 km
s$^{-1}$. Finally, we compare our measurements with those
of the ENEAR catalog \citep{Weg03}, since we use a slightly different procedure to calculate $\sigma_0$. The agreement is good, although the $t$ value indicates that the offset is significant at the 95\% confidence level, but not at 90\% level. It is worth mentioning that our sample includes new spectra with better $S/N$ ratio, as compared to the data presented in \citet{Weg03}.

%
%

%
\section{Absorption Line Measurements}
\label{line_meas}
The formal connection between observational data and stellar
population models was initiated by \citet{Bur84} and \citet{Fab85} who established
the so-called Lick/IDS system, where central bandpass and flanking
continua were defined for several absorption lines. To measure
these indices for galaxies, several procedures and calibrations
are required and in this section we describe them and also apply 
several tests to evaluate their quality. 
Later in this paper, we study the relations of these indices with $\sigma_0$.
We postpone to future papers the determination of the global stellar population parameters such as age, [Z/H] and $\alpha$/Fe \citep{Oga08a}, as well as their radial gradient (Ogando et al. 2008b, in preparation).
\subsection{Indices Definitions}
\label{index_def}

The equivalent widths $W$ were measured for two kinds of features:
atomic absorption lines ($W_a$, expressed in \AA) and molecular
bands ($W_m$, in magnitudes). Their general expressions are given
by the equations below:
\begin{equation}
W_a = \left( 1 - {\int _{\lambda _1}^{\lambda _2} F(\lambda)
d\lambda \over \int _{\lambda _1}^{\lambda _2} C(\lambda) d\lambda
} \right)\left( \lambda _2 - \lambda _1 \right)
\label{eqneqwa}
\end{equation}
\begin{equation}
W_m = -2.5\log {\left({\int _{\lambda _1} ^{\lambda _2} F(\lambda)
d\lambda \over \int _{\lambda _1}^{\lambda _2} C(\lambda) d\lambda
} \right)}
\label{eqneqwm}
\end{equation}
\noindent where $F(\lambda)$ is the line flux, $C(\lambda)$ is the
continuum flux, and $\lambda_1$ and $\lambda_2$ are the wavelengths defining the interval within which $F(\lambda)$ and $C(\lambda)$ are calculated.

Statistical errors in the indices $W_a$ and $W_m$ were calculated,
respectively, according to the following expressions:
\begin{equation}
{{\epsilon_{W_a}} = {{\left[{ {{\left({1\over
{\int_{\lambda_1}^{\lambda_2}}C(\lambda)d\lambda}\right)}^2}{{\epsilon_F}^2}}
+ {{\left({{{\int_{\lambda_1}^{\lambda_2}}F(\lambda)d\lambda}
\over {{({\int_{\lambda_1}^{\lambda_2}}C(\lambda)d\lambda})^2}}
\right)}^2}{{\epsilon_C}^2} \right]}^{1\over 2}}{(\lambda_2 -
\lambda_1)}} \label{erroatom}
\end{equation}
\begin{equation}
{{\epsilon_{W_m}} = {2.5}{{\left[{ {{\left({1\over
{\int_{\lambda_1}^{\lambda_2}}C(\lambda)d\lambda}\right)}^2}
{{\epsilon_C}^2}} + {{\left({1\over
{{{\int_{\lambda_1}^{\lambda_2}}F(\lambda)d\lambda}}}
\right)}^2}{{\epsilon_F}^2} \right]}^{1\over 2}}} \label{erromol}
\end{equation}
\noindent where $\epsilon_F$ and $\epsilon_C$ are the $rms$ values
for the central and continuum features of the line measured. We
have determined, for each galaxy, all the set of Lick indices within the observed spectral range. 
However, to define the final set of indices considered we took into account two conditions: a common spectral coverage to all observing setups and the variation of the fractional error of a given index with $S/N$. The different sensitivity functions produced by the combination of CCD and grating resulted in better determination of some indices than others. 
In particular, the blue indices measurements are poor, as they are in a region of low detector and grating efficiency. 
In Table~\ref{tab:line_def} we list the indices measured with their
passband and pseudo-continua definitions. 
We also include the [O{\sc{iii}}]$\lambda$5007 line definition as given by
\citet{Gon93}, although it is not defined in the Lick system. 
This index is used to identify and discard undesired emission contaminated spectra, as discussed below.

\begin{figure}
  \includegraphics[angle=-90,scale=0.5]{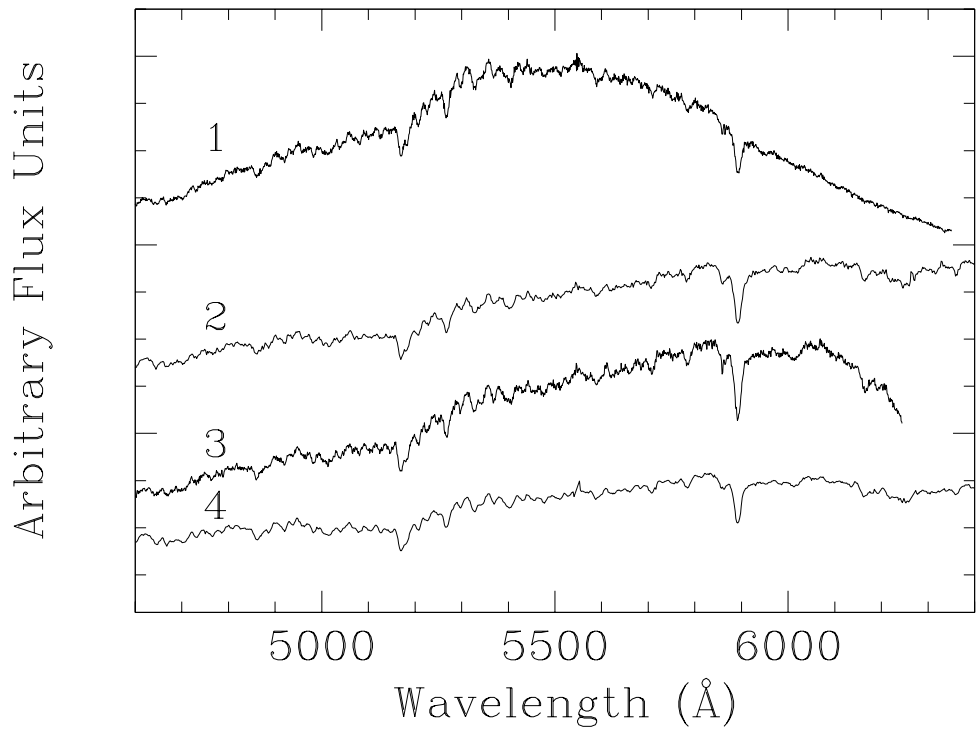}
  \vspace{0.0 cm}
  \caption{The galaxy NGC 7507 is taken as an
example of the typical spectra in each of the setups mentioned in
Table~\ref{tab:line_def}. The setup identification number is shown
at the left side of each spectrum. No flux calibration was
applied. }
  \label{n7507}
\end{figure}

In Figure~\ref{n7507} we show, as an example, the spectra for NGC 7507 as observed with each instrumental setup. Note that the spectrum shape in configuration 1 is quite different from the others. In order to verify the impact that this may have using spectra that have not been flux calibrated we have used a sub-sample of galaxies for which flux calibration is available. We have then compared the line index measurements carried out in both subsets. We find that for the molecular indices Mg$_1$ and Mg$_2$ there is indeed a significant difference (as estimated from a Student $t$ test). This is not unexpected since the continua definition for these indices is far from the central passband.
In the next section we discuss this issue and the procedure to remove this effect.

\subsection{Indices Calibrations}
\label{index_cal}
\subsubsection{Correction to the Lick Resolution} \label{cor_lick_res}

\begin{figure}
   \includegraphics[angle=-90,scale=0.5]{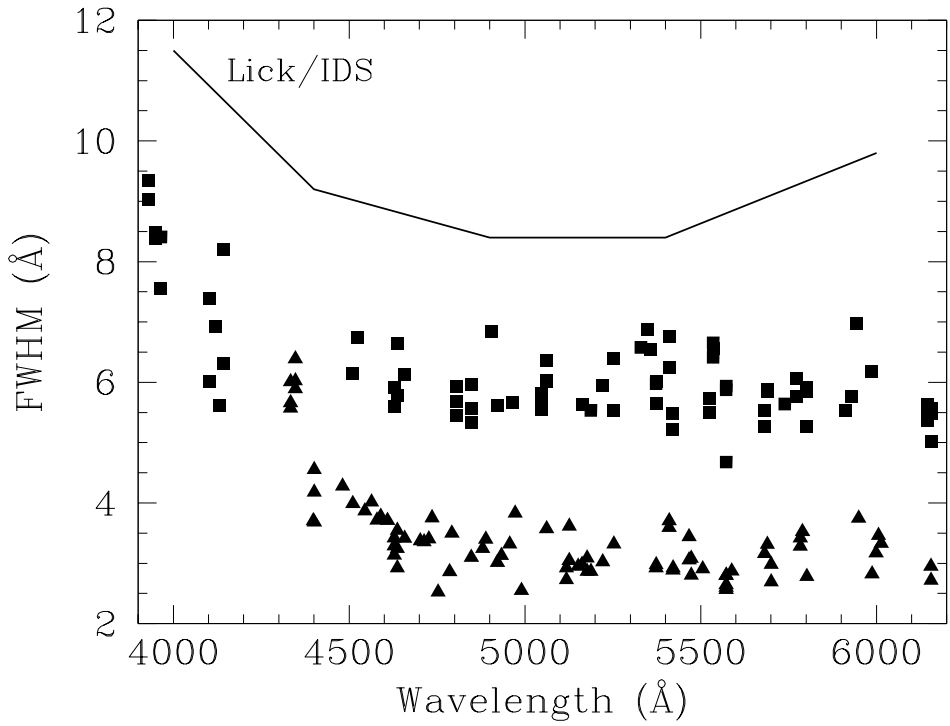}
   \vspace{0.0 cm}
   \caption{Measurements of the FWHM for lines in arc lamp spectra for 1200 l mm$^{-1}$
   (triangles) and 600 l mm$^{-1}$ (squares) gratings. It is also shown the Lick/IDS
   resolution as given by \citet{Wor97} (continuous line). }
   \label{res_redes}
\end{figure}

Since we have a higher resolution than that of the Lick/IDS
system, before calculating indices, we degrade our spectra in
resolution to match that of the Lick/IDS system by convolving the
spectra with a gaussian function with an appropriate width, as
described by \citet{Wor97}. In order to determine our resolution
we fitted gaussian functions to lines of the wavelength
calibration arc lamps. This procedure allowed us to estimate the
resolution variation along the entire wavelength range. 
Figure \ref{res_redes} shows the measured FWHM for several lines
for the two gratings used. For comparison, we plot the Lick
resolution curve described by \citet{Wor97}. It is
important to consider this effect since some of the indices (e.g.:
Mg$b$, Fe5270, Fe5335) are very sensitive to resolution. An
\texttt{IRAF} task named \texttt{lickeqv} was created to carry out
the degradation and also measure equivalent widths.

Despite the conversion of our observed spectra to the Lick/IDS
resolution, this is not enough to fully transform our data to that
system. Small zero-point residuals still persisted. Offsets were,
then calculated for each of the instrumental setups described in
Table~\ref{tab:setups} using the same group of 21 stars of
spectral types F to K and luminosity classes from I to V. This is
true except for setup 1, which had only 9 stars in common with the
others setups. In this case, we estimated the offset relative to setups 2 and 3,
which had already been corrected to the Lick system. 
The choice of a homogeneous set of stars is important
since variations in the offset amongst setups are produced by the
use of different samples of stars. This is probably associated to
internal systematic uncertainties in the Lick system. Spectra of M
stars were available but were not included because of their
non-monotonic behavior with resolution degradation. 

The offsets computed as described above were applied to the indices calculated for galaxies, after the corrections described in the next two sections, as a last step to perform the transformation to the Lick system. Afterwards, as a
final verification, we looked for any residual offset in the
galaxies line strength measurements between setups. Only Mg$_1$
and Mg$_2$ have shown a significant offset between setup 1 and the
others. As mentioned in \S\ \ref{index_cal}, this is probably
related to the lack of flux calibration. This offset was taken
into account and the results show that we ended up with a
homogenous system of line strength measurements that is
appropriate for stellar population analysis.

\subsubsection{Correction for Velocity Dispersions}
\label{cor_vel_disp}

The stellar velocity dispersion in a galaxy causes a spillover of
line flux outside the narrow central bandpass, as it was defined
originally for stars. To compensate this effect we measured line
strengths in artificially broadened spectra of G and K giants. We
simulate velocity dispersions in the range 50-500 km s$^{-1}$ with
steps of 30 km s$^{-1}$. This was done using a modified version of
the \texttt{gauss} task of \texttt{IRAF} package, where the
gaussian dispersion is expressed in pixels, so that the
transformation to km s$^{-1}$ can be done by the relation:
\begin{equation}
\sigma(\mathrm{km~ s^{-1}})=c\frac{\delta\lambda}{\lambda}\sigma(\mathrm{pixels})
\end{equation}
\noindent where $c$ is the velocity of light, and $\lambda$ is a
reference wavelength (5400 \AA). The values of the spectral
dispersions are $\sim$1 \AA\ and $\sim$2 \AA\ for the 1200 l
mm$^{-1}$ and 600 l mm$^{-1}$ gratings, respectively.

\begin{figure}
   \includegraphics[angle=0,scale=0.5]{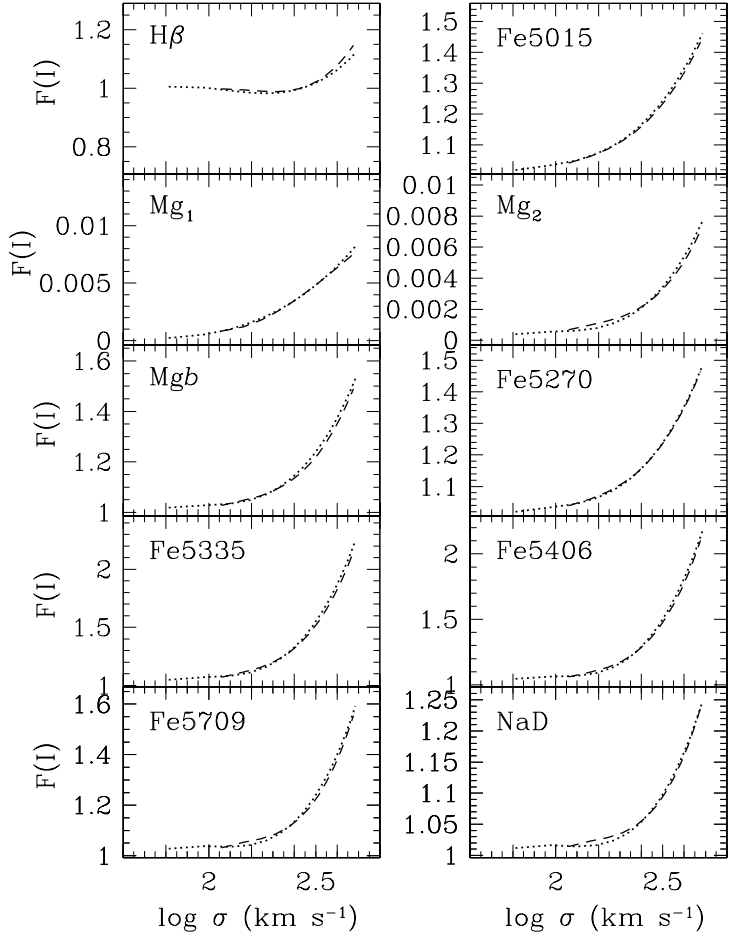}
   \vspace{0.0 cm}
   \caption{Correction factor F(I) for the velocity dispersion. We show the $3^{rd}$ order fits for the grating of 1200 l mm$^{-1}$ (dotted line) and 600 l mm$^{-1}$ (dashed line). }
   \label{factor}
\end{figure}

Measurements of indices in stellar spectra artificially broadened
permit the calibration of a relation between the broadening effect
on the indices, F(I), and $\sigma$. For atomic indices the
relation is F(I)=I(0)/I($\sigma$), while for molecular indices it
is F(I)=I(0)-I($\sigma$), where I($\sigma$) is the index strength
at a given $\sigma$ and I(0) is the index measured in the original
spectrum. Functions F(I) for each index and grating were generated
from 3$^{rd}$ order polynomial fits to these relations for a group
of stars of types G to K according to the expression:
\begin{equation}
\mathrm{F(I)}=\sum^3_{i=0}A_i(\log\sigma)^i
\end{equation}
\noindent In Figure~\ref{factor} we display these functions for
each measured index for both gratings of 1200 l mm$^{-1}$ and
600 l mm$^{-1}$.
Qualitatively, this correction agrees well with others presented
in the literature \citep[e.g.,][]{Tra98,Kun00}. For example, the
amount of correction is very small for broad indices like Mg$_2$,
but large for narrow indices like Fe5270 and Fe5335, reaching
$\approx$ 30\% for galaxies with 300 km s$^{-1}$. Another point is
that the H$\beta$ index is very sensitive to the choice of stellar
spectral type into the definition of F(I), as already pointed out
by \citet{Kun00}, producing sometimes values of F(I) smaller than
1. The causes for this behavior are that the H$\beta$ itself fades
very quickly as the temperature gets lower, and a TiO band head
($\approx$4851\AA) starts to appear in low temperatures stars
\citep{Schi02i}. For this reason, in the particular case of the
H$\beta$ index, some of the lowest temperature stars are not
included in the fit. Quantitatively, the average
difference of F(I) at the arbitrary value of $\sigma=300$ km
s$^{-1}$ is about 1$\pm$3\% for the indices in common with
\citet{Tra98}, with the exception of Mg$_1$ and Mg$_2$. For the Mg
indices they use a multiplicative factor instead of an additive one,
thus not allowing a comparison.

\subsubsection{Correction by Aperture}
\label{cor_aperture}

The indices calculated as described above do not refer to a fixed
metric aperture or to some fraction of $r_e$, implying that we
sample differently the galaxies, as described in \S\ \ref{redu}.
Since the distribution of chemical elements inside the galaxy is
not homogeneous, we need to correct them to some fiducial aperture
in a similar way to what was done for $\sigma_0$ in \S\ \ref{mea_vel_disp}.
Here, we also consider an average gradient, this time for line strengths,
which are all measured in magnitudes, even in the case of atomic indices,
so that they are converted from \AA\ (I) to magnitudes (I$'$) according
to the equation:
\begin{equation}
\mathrm I'=-2.5\log\left(1-\frac{\mathrm I}{\Delta\lambda}\right)
\label{conv_ind_mag}
\end{equation}
\noindent where $\Delta\lambda$ is the central passband of the
index. Then we derived a fiducial aperture corrected index
(I$'_{norm}$) given by:
\begin{equation}
{\mathrm I}'_{norm}={\mathrm I}'_{ab}-\beta\log\left(\frac{r_{ab}}{r_{norm}}\right)
\end{equation}
\noindent where $r_{ab}$ and $r_{norm}$ have the same meaning as
in equation (\ref{sigma_cor_Jor}), and $\beta$ is the average radial
gradient for each index, which is presented in
Table~\ref{tab:med_gradind}. The parameter $\beta$ is the angular
coefficient of a straight line fitted to the I$'(r)$ profile as a
function of $\log(r/r^*_e)$, where $r^*_e$ is the galaxy's
effective radius corrected for its ellipticity. A full description
of this parameter and its determination will be presented in
Ogando et al. 2008b. 

\subsubsection{Emission Contamination}
\label{cor_emission}

Early-type galaxies may contain some amount of ionized gas and
young stars, which may lead to the presence of emission lines and
contamination of their old population spectra
\citep[e.g.,][]{Mac96}. In these cases, some line indices may be
significantly affected by these emission features, in particular
H$\beta$ and Fe5015 (by the [O {\sc{iii}}]$\lambda$5007 line).
Indeed, even if the emission is not strong, it may fill partially
the H$\beta$ line, resulting in a smaller value for its measured
equivalent width and as a consequence, causing ages to be overestimated when using SPP models \citep{Tra00}. A possible way to correct for the
H$\beta$ contamination can be accomplished if we can infer the
contribution from another emission line. For the case of H$\beta$,
\citet{Gon93} derived a relation between the emission
contamination in this line and the [O{\sc{iii}}]$\lambda$5007
equivalent width, established by the emission-free stellar spectra
fit to those galaxies in his sample. He found that the amount of
emission contamination in H$\beta$ is given by
$\Delta$H$\beta=0.6\times$[O{\sc{iii}}]$\lambda$5007. However this
correction has a considerable uncertainty and should be taken as a
statistical approach since the H$\beta$/[O{\sc{iii}}]$\lambda$5007
ratio varies considerably from galaxy to galaxy.

\begin{figure}
   \includegraphics[angle=0,scale=0.6]{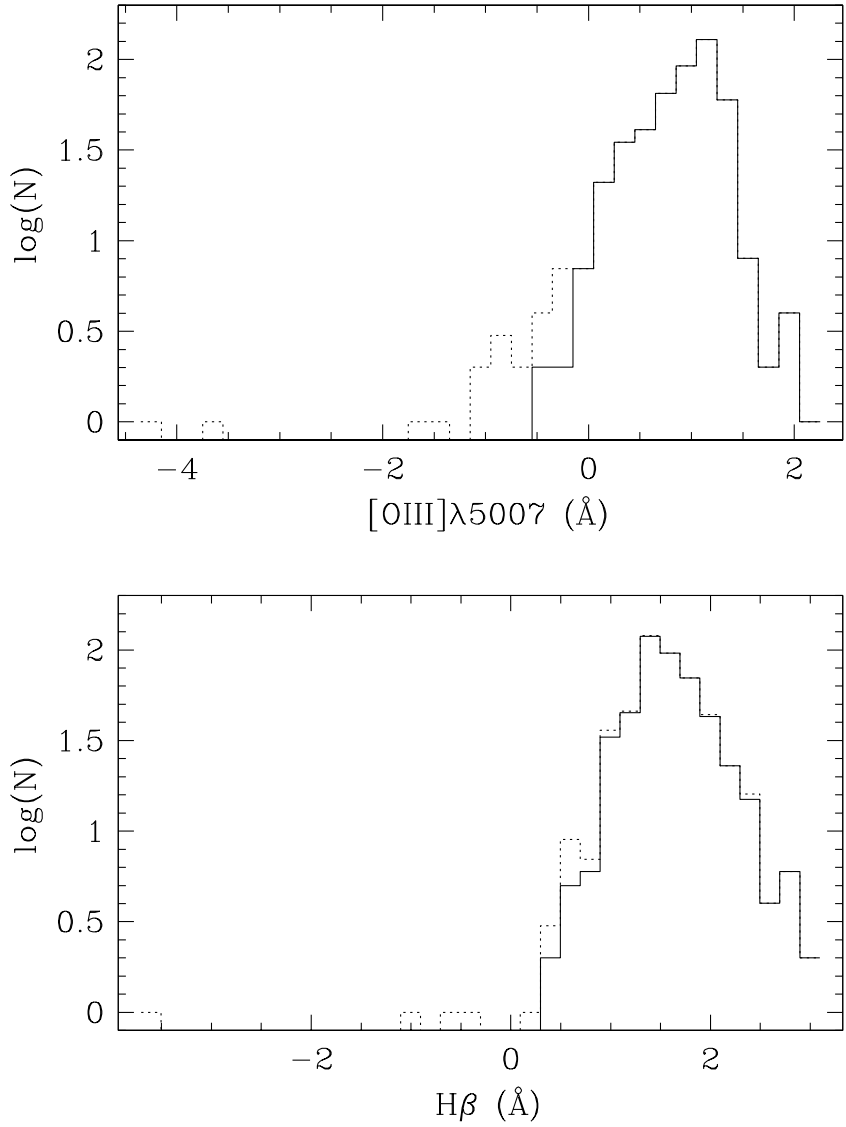}
   \vspace{0.0 cm}
   \caption{Distribution of $[$O{\sc iii}$]\lambda 5007$ (top panel) and H$\beta$
   (bottom panel) before (dotted line) and after (solid line) discarding galaxies
   with detected emission.  }
   \label{emhist}
\end{figure}

Another possible way to deal with the emission contamination in
the H$\beta$ index is to infer it through the intensity of
H$\alpha$ line \citep[e.g.,][]{Nel05,Den05a}. However, due to the
different spectral coverage associated to the several instrumental
setups used, we are able to measure H$\alpha$ for only one third of the
sample. Thus, instead of using these statistical corrections, we
decided to exclude from the analysis galaxies with measurable
emission of [O{\sc{iii}}]$\lambda$5007 at a detection level larger
than twice its rms error. This condition affects 18 galaxies, or
about 4\% of the sample, which is a smaller fraction than that of
12\% found by \citet{Nel05} for galaxies selected on the basis of
red sequence criteria which may include some early-type spirals,
since no explicit morphological segregation was imposed. The
galaxies with detected emission are predominantly S0 (11
galaxies), and they are distributed evenly among the distinct
environments. We illustrate the impact of this rejection criteria
on the distribution of $[$O{\sc{iii}}$]\lambda 5007$ and H$\beta$
intensities in Figure \ref{emhist}, where we note that even
galaxies with positive H$\beta$ indices are excluded from the
sample, showing how subtle the emission contamination can be.

\subsection{Internal Comparison}
\label{comp_ourselves}

For objects with several observations, the final value for an index was computed by: 1) applying to each available measurement all the calibrations and corrections mentioned above; 2) computing the mean of all measurements: 3) discarding outliers with values 2$\sigma$ above the average; 4) re-computing the mean one more time.
The final values for the indices of all the 509
galaxies are presented in Table \ref{tab:line_indices}, including
those with emission lines which are not included in the analysis
below.

\begin{figure}
   \includegraphics[angle=0,scale=0.6]{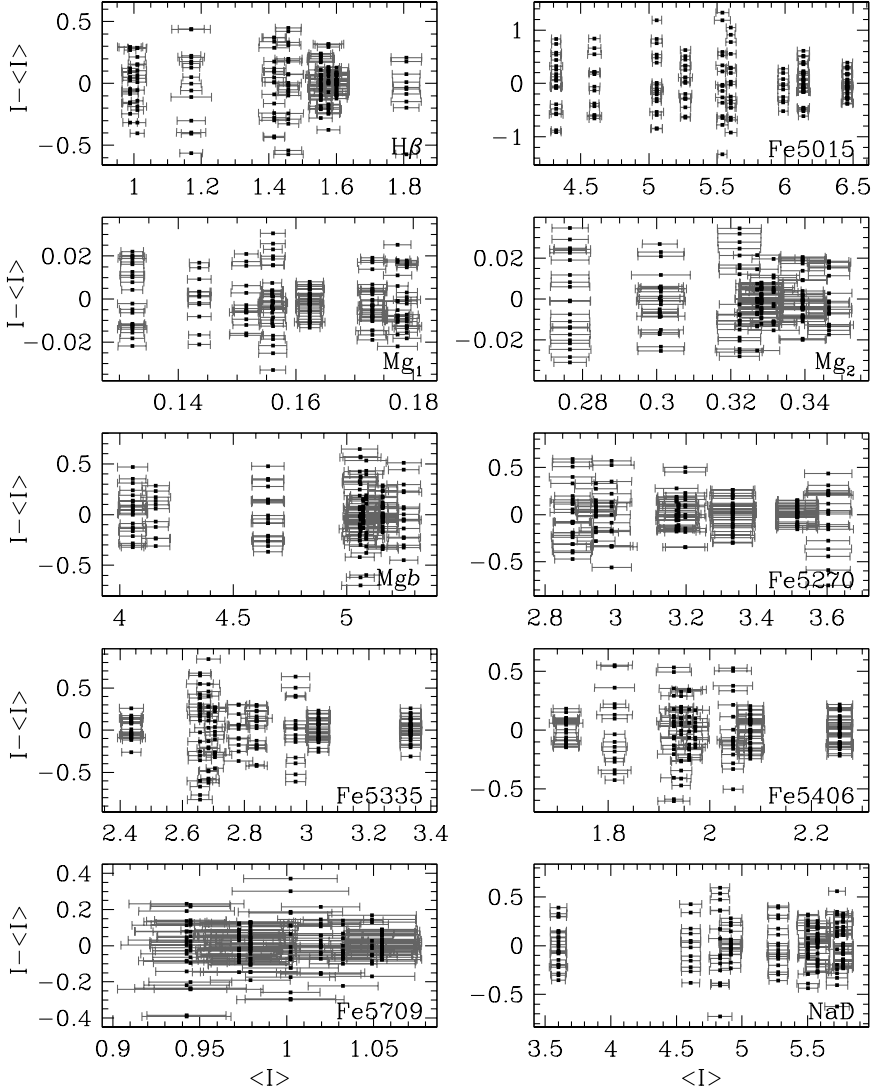}
   \vspace{0.0 cm}
   \caption{Scattering of line indices for galaxies with more than
   ten observations. The panels show the fractional errors ($\rm I-\langle I\rangle$) for
   the various line indices versus the average line index ($\langle\rm I\rangle$). The
   horizontal error bars indicate the error of each individual measure.
   }
   \label{comp_entrenos}
\end{figure}

Besides the statistical error determination described
in \S\ \ref{index_def} we need to take into account other
sources of uncertainties as, for example, those related to
wavelength calibration, sky background subtraction, position angle
of the slit over the galaxy, the fraction of galactocentric radius
encompassed by the slit. Multiple observations ($N>2$) are
available for 91 galaxies ($\approx18$\%) of our sample, and they
are used to estimate a more representative error for our line
strengths. We calculated for each line index an average of the
fractional error combining the individual determinations. In Table
\ref{tab:statmultobs} we present the statistics of this averaging
process and it can be noted that, in general, the errors derived
from multiple observations reach 5-10\% of the index value, while
errors from individual measurements are typically 1-3\% of the
index value. In Figure \ref{comp_entrenos} we display measurements
of indices for galaxies with more than 10 observations
in order to show the range of their fluctuations.
Thus, based on the statistics shown in Table \ref{tab:statmultobs}
we claim that, in general, our errors are less than 10\% of the index value.

\subsection{Comparison with other Authors}
\label{comp_outros_autores}

\begin{figure}
   \includegraphics[angle=0,scale=0.6]{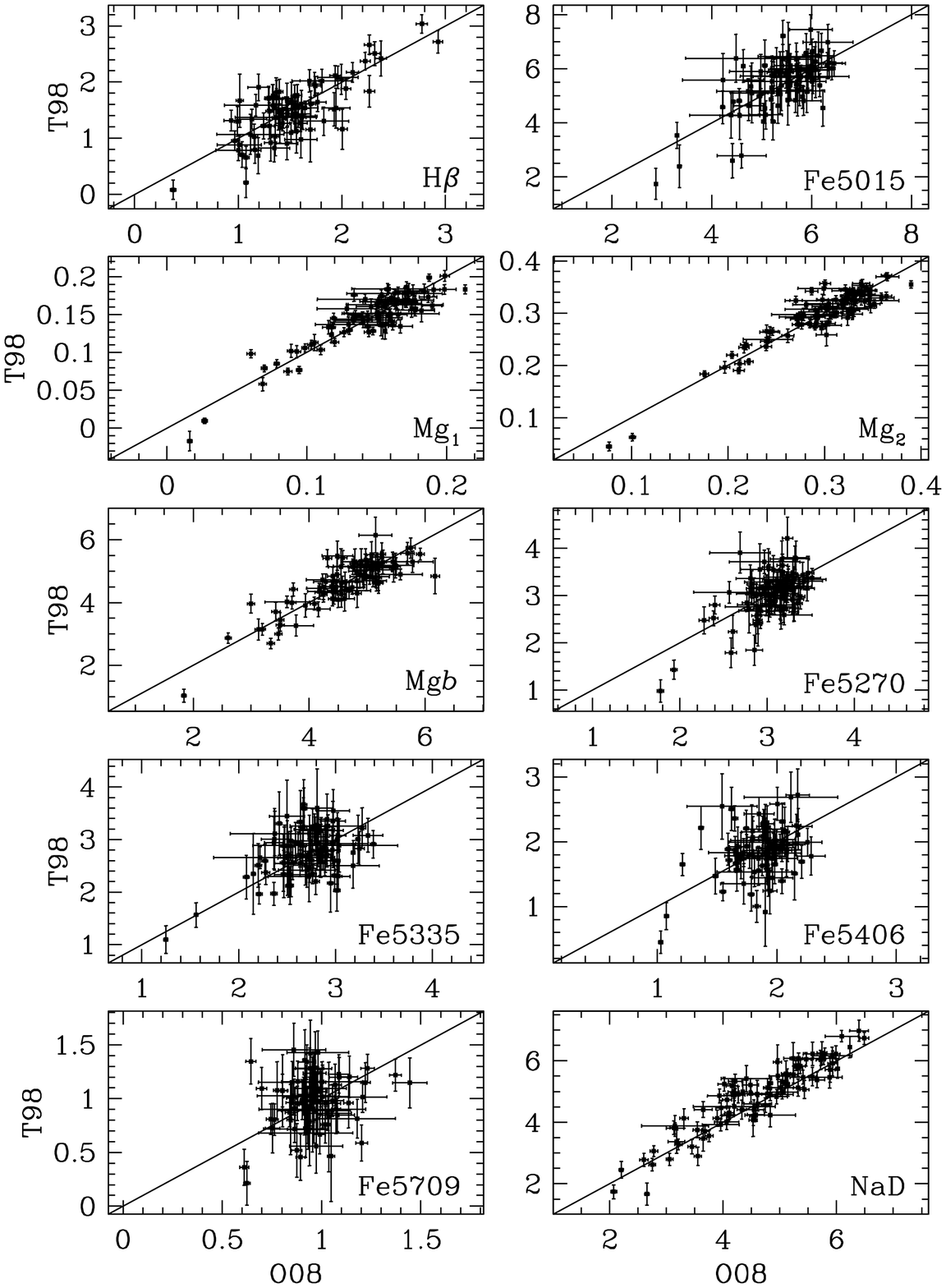}
\vspace{0.0 cm}
   \caption{Comparison between galaxies indices measured by \citet{Tra98} and our
   results (O08).
   }
   \label{comp_indices_1}
\end{figure}

\begin{figure}
   \includegraphics[angle=0,scale=0.6]{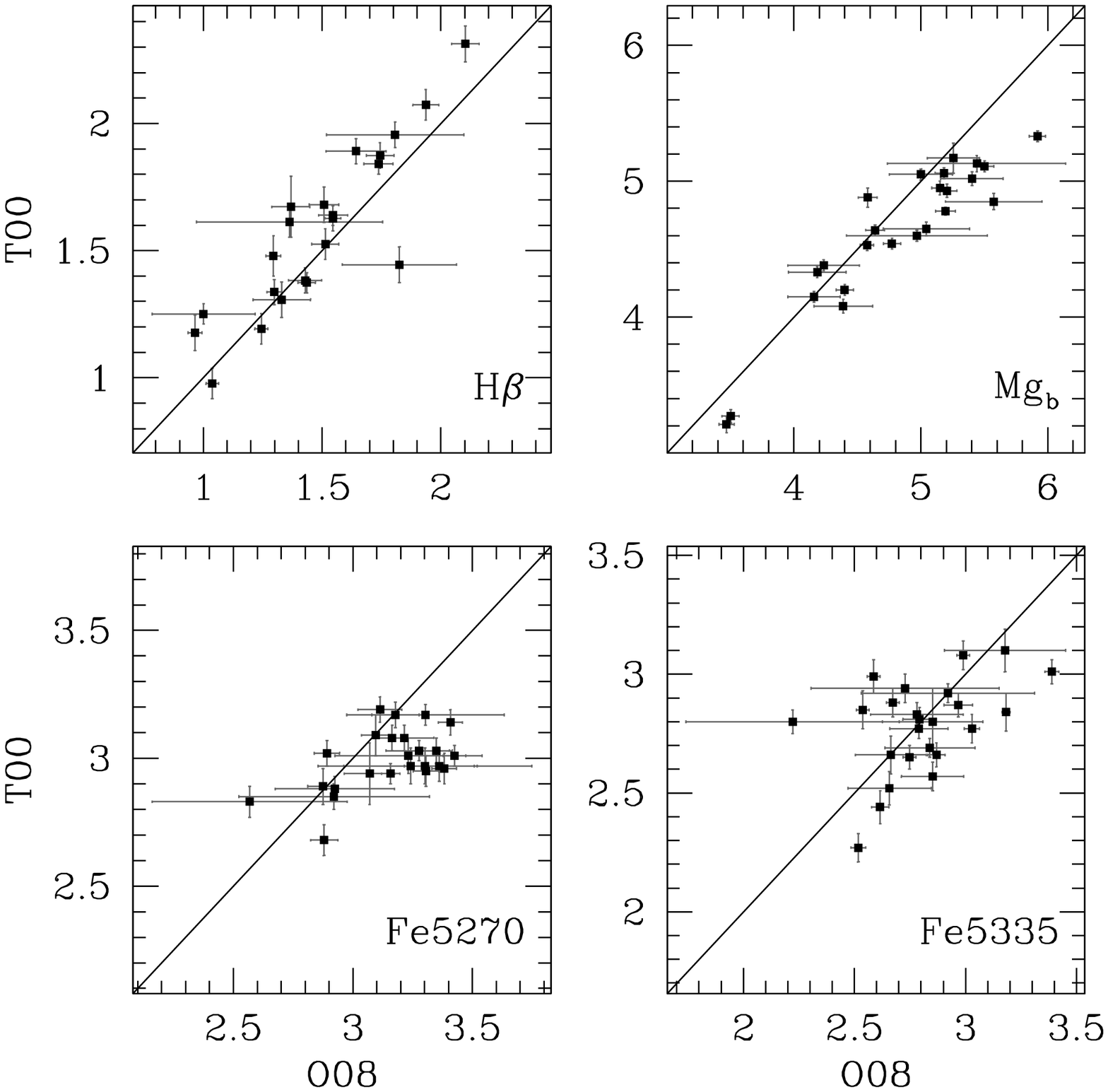}
\vspace{0.0 cm}
   \caption{Comparison between galaxies indices measured by \citet{Tra00} and our results (O08).
   }
   \label{comp_indices_2}
\end{figure}

\begin{figure}
   \includegraphics[angle=0,scale=0.6]{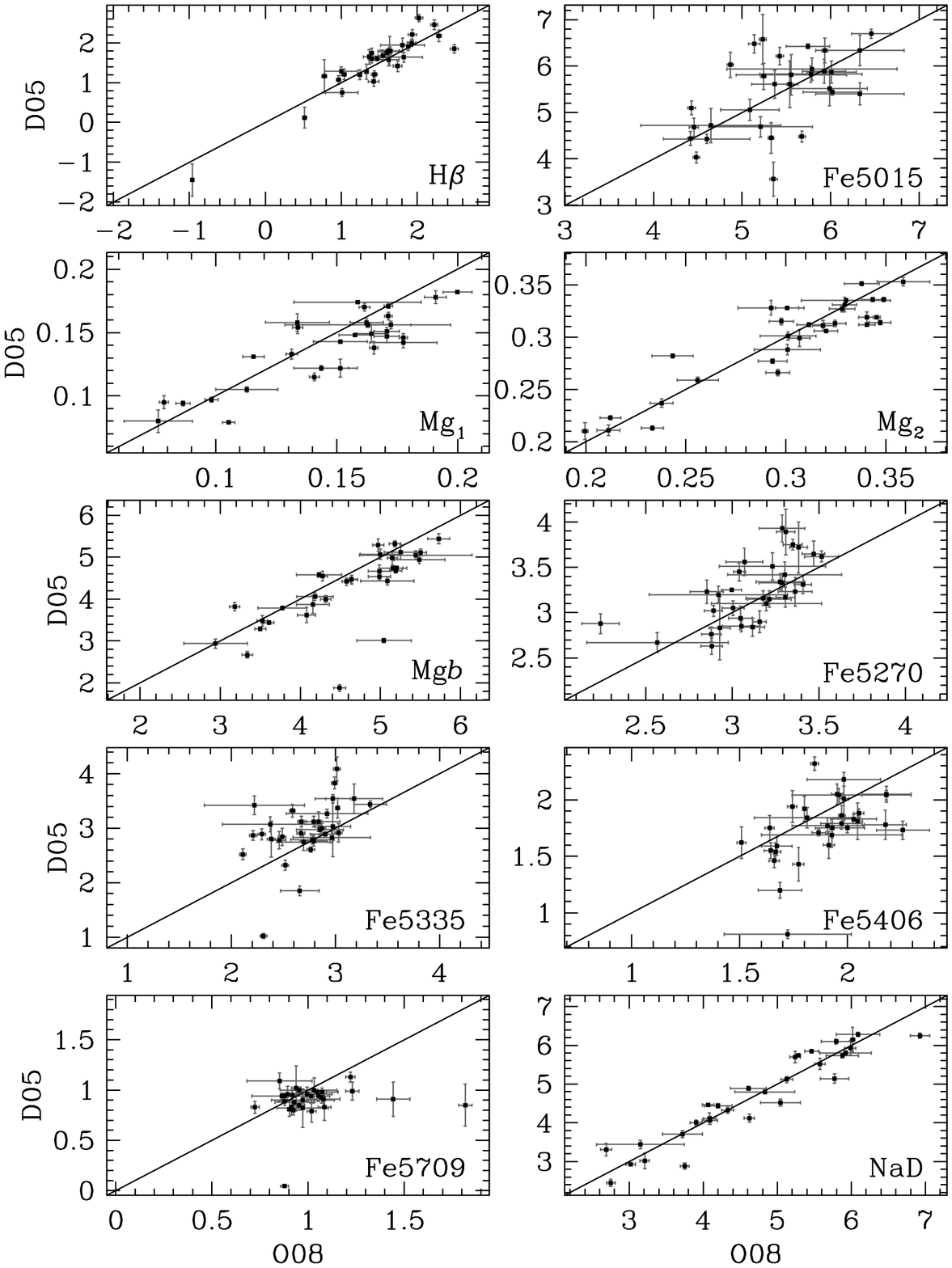}
\vspace{0.0 cm}
   \caption{Comparison between galaxies indices measured by \citet{Den05a} and our results (O08).
   }
   \label{comp_indices_3}
\end{figure}

\begin{figure}
   \includegraphics[angle=0,scale=0.6]{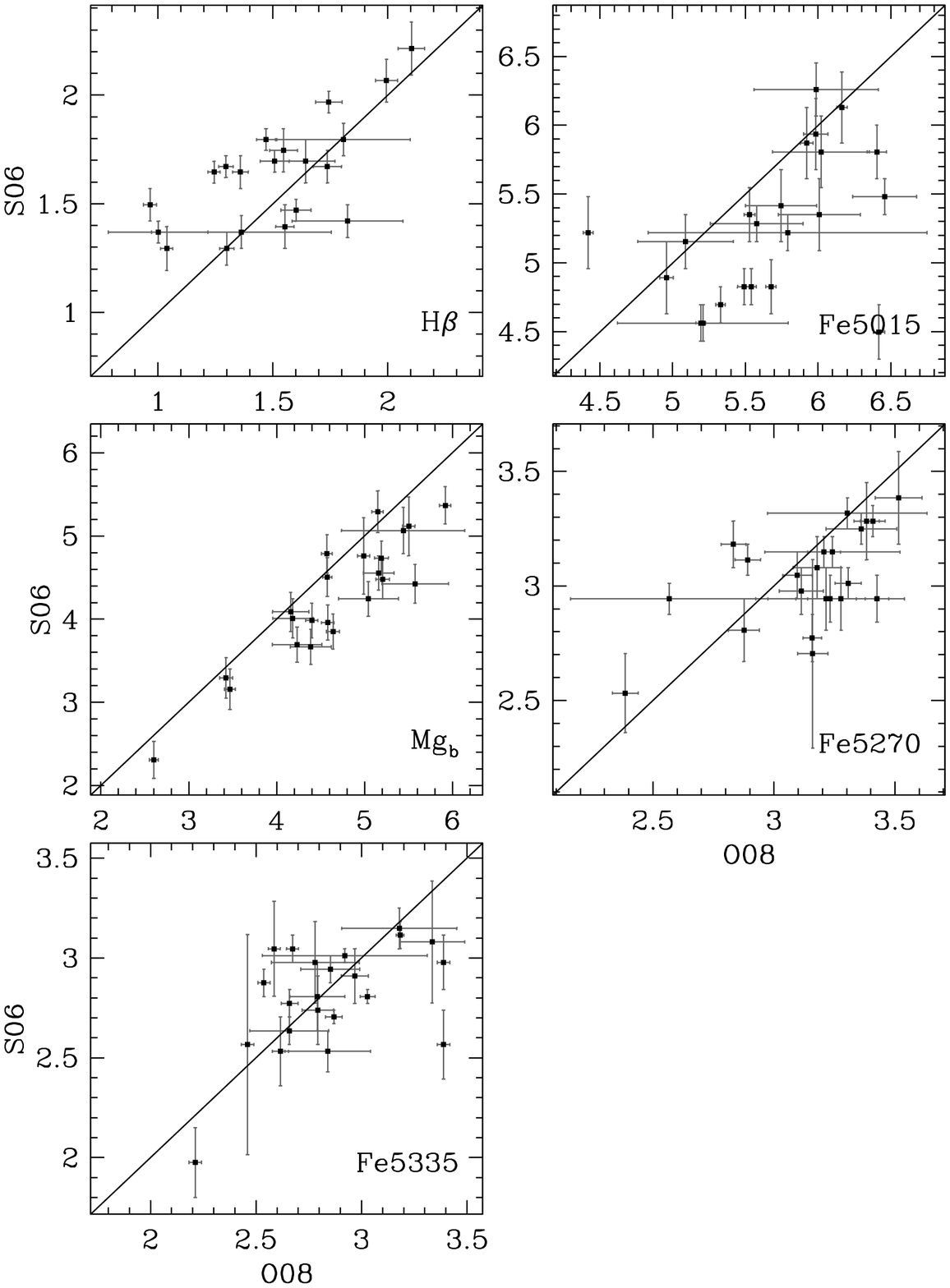}
\vspace{0.0 cm}
   \caption{Comparison between galaxies indices measured by \citet{San06a} and our results (O08).
   }
   \label{comp_indices_4}
\end{figure}

\begin{figure}
   \includegraphics[angle=0,scale=0.6]{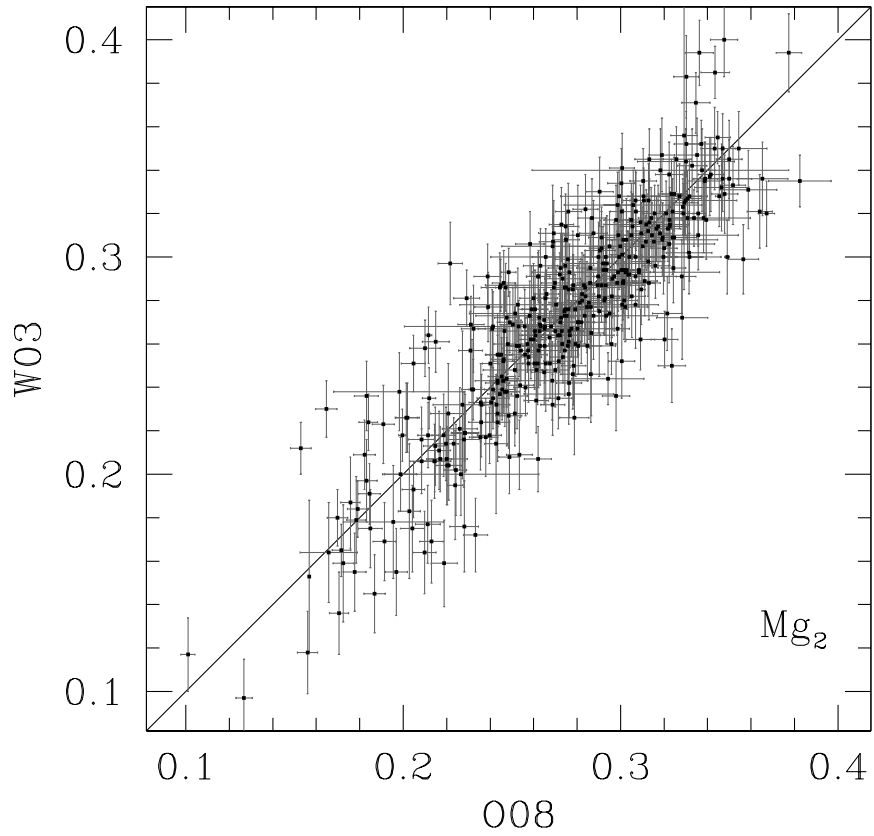}
\vspace{0.0 cm}
   \caption{Comparison between Mg$_2$ index measured by \citet{Weg03} and our results (O08).
   }
   \label{comp_indices_5}
\end{figure}

To evaluate the overall quality of our data and calibrations, we have
compared our measurements of line indices with those from measurements reported in the literature by other authors of samples having a significant number of overlaps with ours. The results of these comparisons are shown in Table \ref{tab:comp_todos_autores} and Figures \ref{comp_indices_1} to \ref{comp_indices_5}. Table  \ref{tab:comp_todos_autores} gives: in column (1) the reference; in column (2) the number of galaxies in common $N_{gal}$; in column (3) the mean difference $\langle\delta\mathrm I\rangle$ and its $rms$; and in column (4) the $t$ value of the Student test. We remind that for $t>1.96$, the differences are considered significant at a 95\% confidence level.

Figure \ref{comp_indices_1} shows the comparison with \citet{Tra98}, which includes the original Lick/IDS data. Despite of the lower quality of the data, we find from Table \ref{tab:comp_todos_autores} that the agreement is very good. The same is true regarding the comparison with \citet{Tra00} (Figure \ref{comp_indices_2}) which reports data of better quality obtained by \citet{Gon93}. The only exception is the Fe5270 index. More recently, new data have been presented by \citet{Den05a} and \citet{San06a}. Comparison with their measurements are shown in Figures \ref{comp_indices_3} and \ref{comp_indices_4}. We find that in general the agreement is good, except for the Fe indices measured by \citet{Den05a}. In the case of \citet{San06a} only the Fe5015 presents
a more discrepant result. Finally, in Figure \ref{comp_indices_5} we compare our measurements of the Mg$_2$ index with those presented earlier in \citet{Weg03} for $\approx400$ galaxies in common, but occasionally using new spectra with better $S/N$. We note that in the present paper we use a slightly different method to estimate the pseudo-continuum on each side of the band trying to correct for the local inclination of the spectra. The quantitative results are given in 
Table \ref{tab:comp_todos_autores}, from where we find that the agreement is excellent.

We also mention that in the last years large surveys similar to the one being reported here have been conducted \citep[e.g.,][]{Nel05,Col01} but unfortunately the number of objects in common is very small ($< 3$) to allow a comparison.

In summary, based on the results presented above, we conclude
that, in general, our measurements of line index are in good agreement with those available in the literature. Occasionally, some significant discrepancies are detected for a particular index, but this does not seem to reflect a systematic problem as it varies from author to author. Instead, they probably reflect different effects that may affect the measurement such as differences in the region of the galaxy sampled (e.g. different position angle and/or radial extent), lack of a suitable set of stars used to calibrate the Lick system by some authors, and  the details of the procedure adopted for estimating the correction due to velocity dispersion and resolution.

%
\section{Indices, Mass, Velocity Dispersion and Environment}
\label{index_sigma}

A connection between dynamical and chemical properties of
early-type galaxies may be established through relations between
the strength of line indices and their velocity dispersions or
masses, as for example the well known Mg$_2-\sigma_0$ relation.
Since our data constitute one of the largest samples of
early-type galaxies with good spectral quality and resolution
at $z\approx 0$ we investigate not only this but similar relations for
other lines taking into account different morphologies and
environments. In order to do that, we first define environment and
mass estimators for our galaxies as described below.

\subsection{Characterizing the Galaxy Environment}
\label{gal_env}

Although early-type galaxies are found more frequently in clusters
and groups of galaxies, their relative quantity to other
morphological types depend essentially on the local density of
galaxies. This behavior is translated in the well known
``morphology-density'' relation \citep[e.g.,][]{Dre80, Mai90}.
Taking this into account, we decided to characterize the
environment for a galaxy estimating the number of
companions brighter than a fixed absolute magnitude, within a
given ``box'' size. This is equivalent to calculate the density of
objects, being more representative of the local environment around
an object than just classifying it as a field, group and cluster galaxy.

\begin{figure}
   \includegraphics[angle=0,scale=0.6]{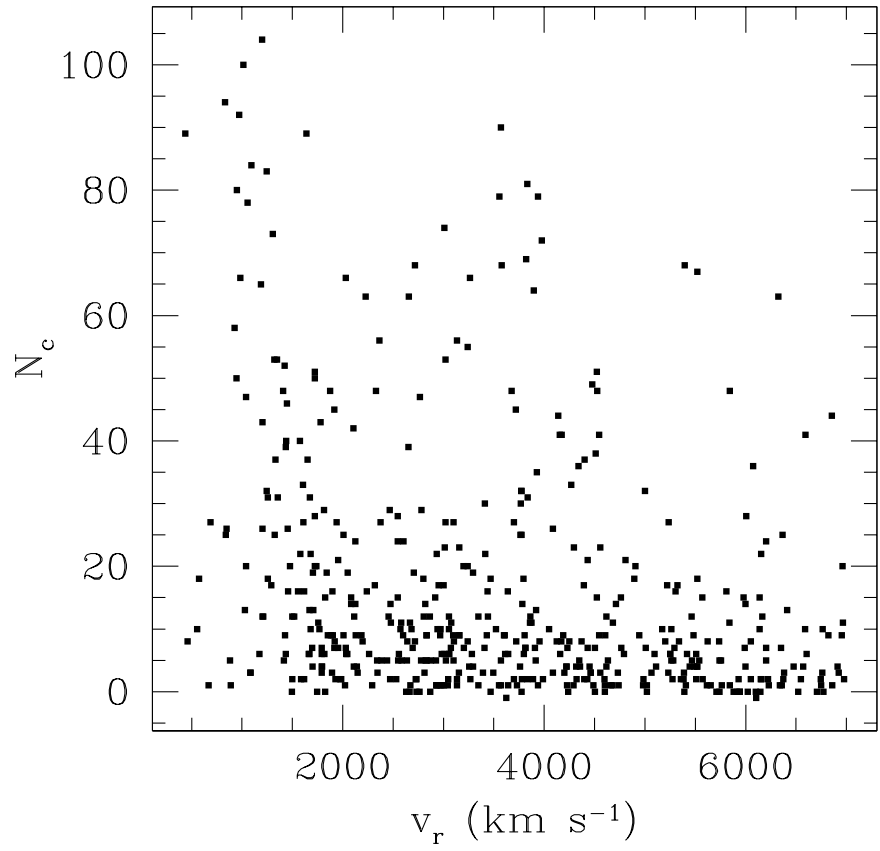}
   \vspace{0.0 cm}
   \caption{Number of companions ($N_c$) for the galaxies of our sample versus the
   radial velocity $v_r$. The concentration of galaxies around $v_r$ 1000 km s$^{-1}$
   is related to the Virgo cluster. }
   \label{namb_vel}
\end{figure}

\begin{figure}
   \includegraphics[angle=-90,scale=0.6]{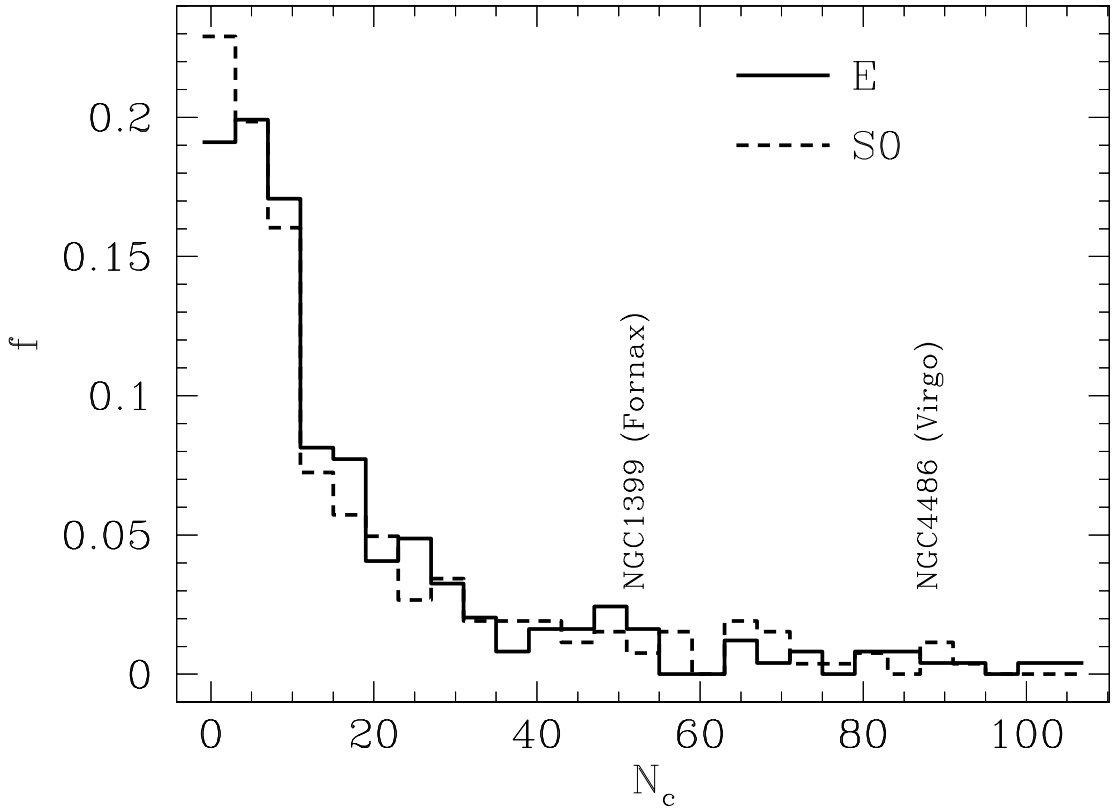}
   \vspace{0.0 cm}
   \caption{Frequency distribution of the number of companions ($N_c$) according to
   morphological type, E (continuous line) and S0 (dashed line). Two galaxies in clusters
   of different richness, NGC 1399 in Fornax and NGC 4486 in Virgo, indicate the sensitivity
   of the $N_c$ determination method to distinct density regimes. }
   \label{num_viz}
\end{figure}

In order to estimate the number of neighbors ($N_c$) for each
galaxy in our sample, we used the catalog of galaxies available in
the HyperLeda database \citep{Pat03}, which has a larger sky coverage
than the Southern Sky Redshift Survey \citep{daC98}, for example.
Although the former represents a compilation, it is considered
complete down to $m_B\approx 15.5$. A galaxy from this catalog
contributes to the neighborhood counts of a given galaxy in our
sample, if it satisfies the following criteria: (i) the difference
between their radial velocities is $<750$ km s$^{-1}$, (ii) their
projected separation is $<500$ kpc, (iii) the HyperLeda galaxy has
$M_B<-16.0$. It should be mentioned that there is not any strong
bias of $N_c$ with distance as can be seen in Figure
\ref{namb_vel}. The distribution of $N_c$ for our sample divided
by morphological type is displayed in Figure \ref{num_viz}. We
note that this density indicator has a wide dynamical range making
it sensitive to variations from the external to central regions of
clusters and to the mean cluster density.
This can be seen in Figure \ref{num_viz} where we indicate in the
plot the position occupied by a galaxy in the central part of the
Fornax cluster (medium density) and from one in the central part
of the Virgo cluster (high density). This indicates that this
classification scheme is more meaningful than the usual
field/group/cluster one, as used by \citet{Ber98}, for instance.

The median value for the $N_c$ distribution in the volume
considered is 10 galaxies and the lower and upper quartiles are
LQ=4 and UQ=22 galaxies, respectively. Taking these values into
account, we divided the sample in 3 intervals of densities: low
density (LD; $N_c\leq4$), median density (MD; $4<N_c<22$), and
high density (HD; $N_c\geq22$). 
Our main goal in splitting the sample in this way is to compare the two extreme environments, LD and HD,
using about the same number of objects in each one of them, so
that we can avoid misleading statistical inferences. Among the 509
galaxies of our sample, only two galaxies were considered isolated
($N_c=0$) according to the prescribed criteria. Examining $N_c$
for 11 galaxies in common with the isolated sample of
\citet{Red04}, only three have $N_c>2$. Among them, NGC 1132, that
is the galaxy with the greatest $N_c$ (11 companions), has an
associated extended x-ray emission typical of groups of galaxies.
Indeed this galaxy is considered a ``fossil group'' by
\citet{Mul99}, and if we inspect DSS images, it is possible to see
fainter galaxies around this object. We also note that among the
criteria adopted by \citet{Red04} there is one that excludes
companions 2 magnitudes fainter than that of the target object,
what could explain this case.

\subsection{Galaxy Mass Estimators}
\label{gal_mass_est}

The mass of a galaxy is one of its fundamental parameters and its
determination may follow different recipes. The most common
estimator is the central velocity dispersion $\sigma_0$ which, in
fact, probes the gravitational potential. Another mass estimator
comes from the consideration that the object is in dynamical
equilibrium, satisfying the virial theorem. Thus, the mass $M_e$
in units of M$_{\odot}$ is given according to \citet{Bur97} by the
expression:
\begin{equation}
\log M_e=\log (\sigma_0^2\ r_e) +5.67
\end{equation}
\noindent where $r_e$ is the galaxy effective radii in units of kpc.

In both cases it is implicit that the contribution by rotation to
the object's dynamical support is assumed to be negligible. Another point of
concern is related to the conversion of $r_e$ in arc seconds to
kpc which needs a reliable determination of the galaxy distance
($D$). Considering just the redshift to calculate $D$, we may
incur in error due to the peculiar motion of galaxies. Thus, we
adopted $D$ given by D$_n-\sigma$ relation \citep{Ber02} to
calculate $r_e$ in kpc.

\begin{figure}
   \includegraphics[angle=0,scale=0.6]{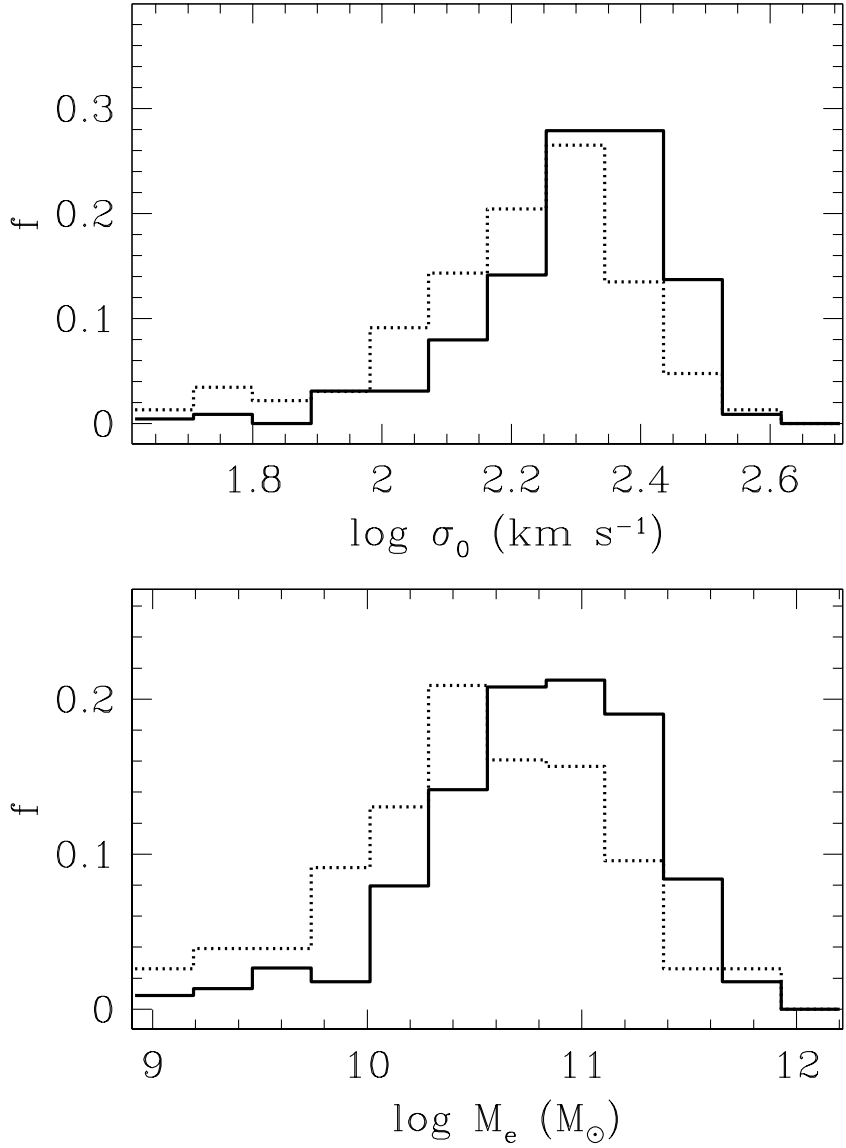}
   \vspace{0.0 cm}
   \caption{Frequency distribution of mass estimators for E (continuous line) and S0
   (dotted lines).}
   \label{sigma_mass_distr}
\end{figure}

Since in the next section we discuss the relations between indices and
mass for the sample divided in different morphologies and
environments, it is interesting to previously inspect the differences
in the mass distribution of these particular
subsets. We show in Figure \ref{sigma_mass_distr} these
distributions using both estimators described above,
according to distinct morphological type. In Table
\ref{tab:mass_gal} we present mean values, dispersion, skewness,
and kurtosis for these estimators. Also, in Table
\ref{tab:cmplvd_gal} we show the probability that the samples of E and S0 galaxies have the same parent population as given by the Kolmogorov-Smirnov
(KS) test. For both mass estimators, these samples have very different distributions, where the mass
distribution of E galaxies is skewed towards higher masses as
compared to the one for S0s. We investigate below if this behavior
implies in different index$-\log\sigma_0$ relations for distinct
morphologies.

\begin{figure}
   \includegraphics[angle=0,scale=0.6]{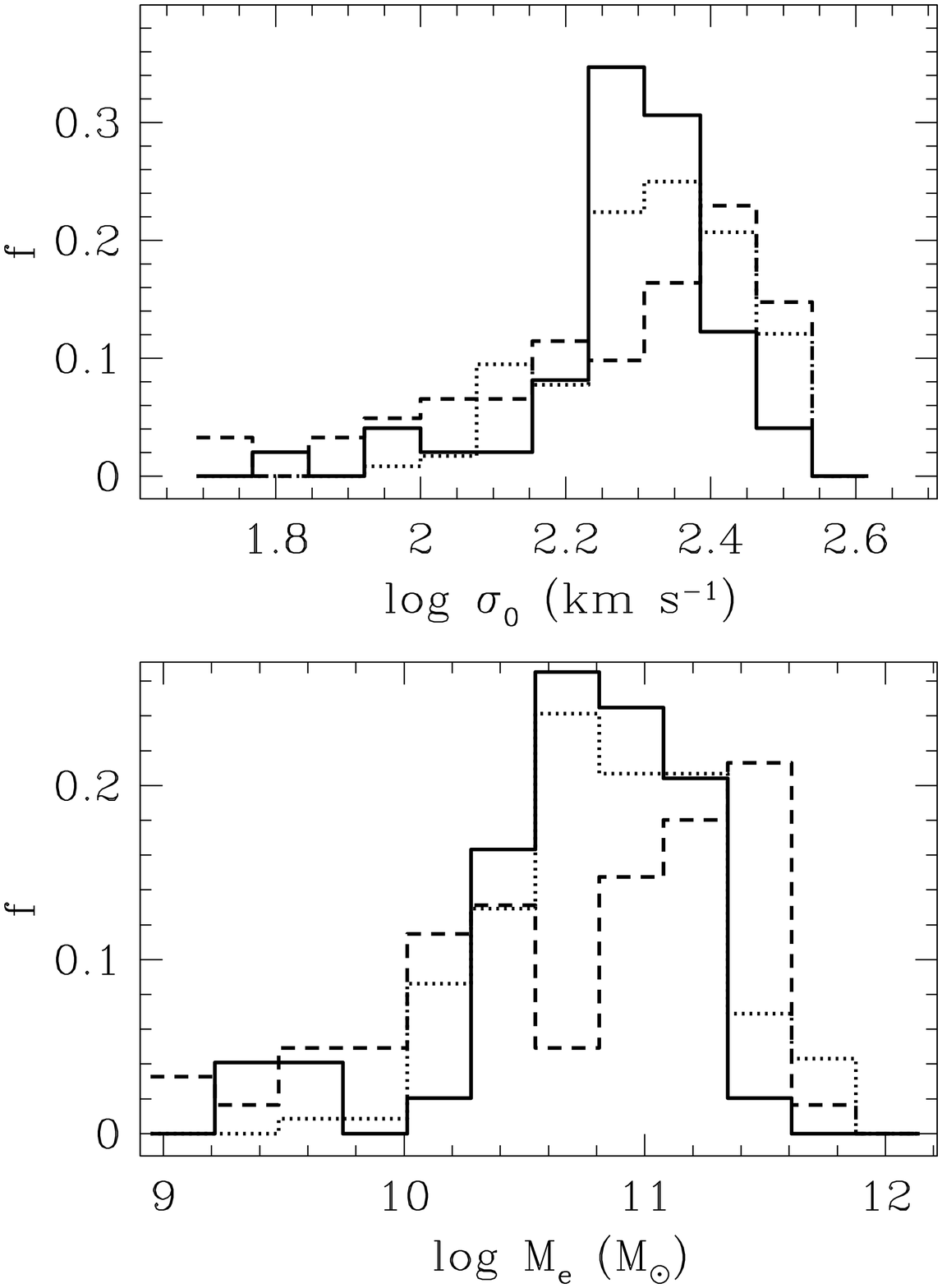}
   \vspace{0.0 cm}
   \caption{Frequency distribution of mass estimators for E galaxies in different
   environments: LD (continuous), MD (dotted), and HD (dashed). }
   \label{dist_mass_e}
\end{figure}

\begin{figure}
   \includegraphics[angle=0,scale=0.6]{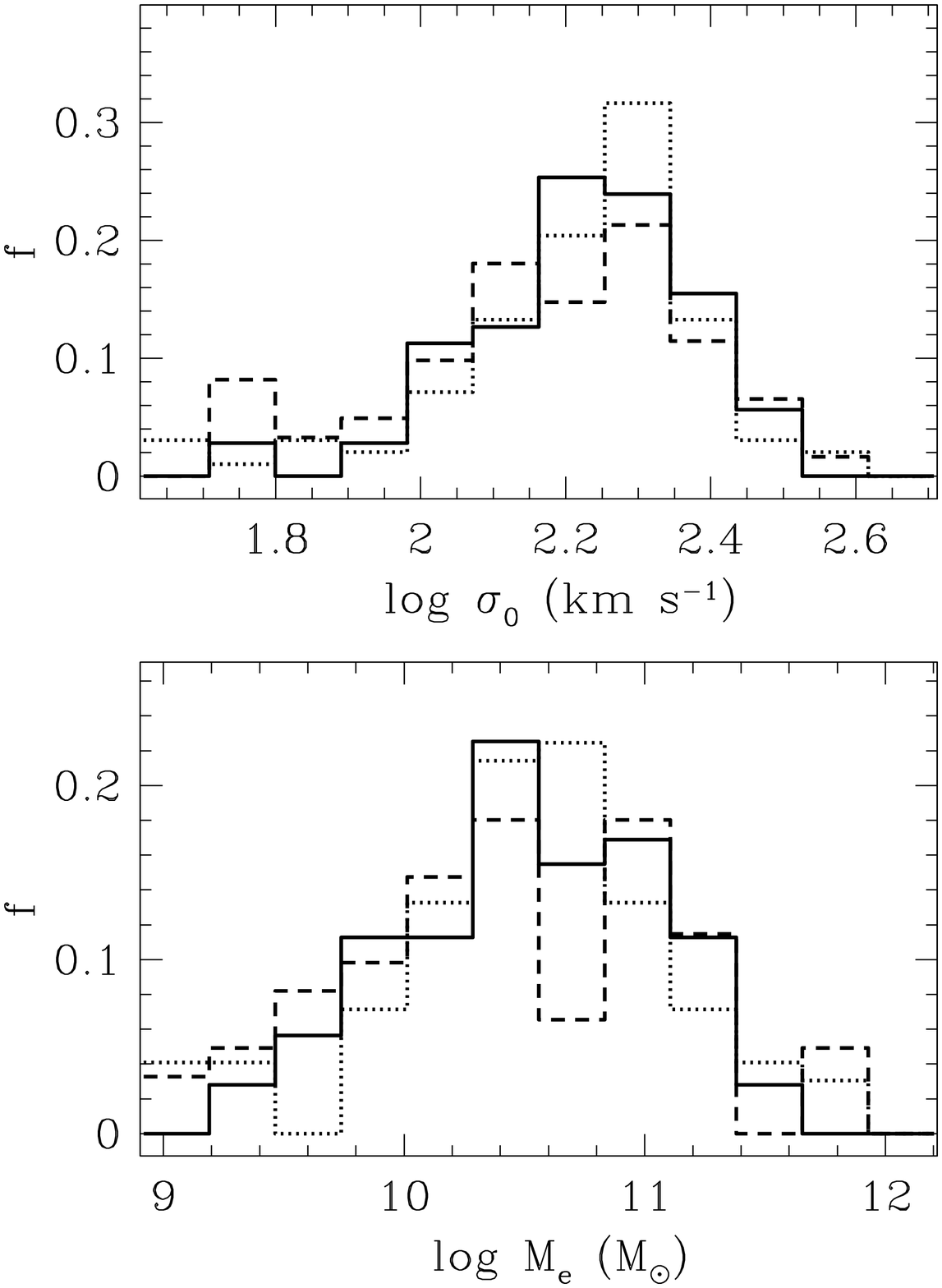}
   \vspace{0.0 cm}
   \caption{Frequency distribution of mass estimators for S0 galaxies in different
   environments: LD (continuous), MD (dotted), and HD (dashed). }
   \label{dist_mass_s0}
\end{figure}

We also use the environment definitions described above to
divide the mass distribution of E and S0 galaxies, as shown respectively in Figures \ref{dist_mass_e} and \ref{dist_mass_s0}.
These plots, which reflect the morphology-density relation, stress
that the most massive Es are located in high density environments,
but S0s do not follow this trend, having similar mass
distribution in all environments.

It is noticeable in Figure \ref{dist_mass_e} and also
from the computed values of kurtosis and skewness in Table \ref{tab:cmplvd_gal}, that the mass distribution of E galaxies in HD is less peaked and more
concentrated towards the high mass end of the distribution than
those in LD environment. For the S0s, such trend occurs more
mildly and, in particular, for $\log M_e$, the similarities
between the shape parameters and the higher KS probability (40\%)
indicate that S0s have almost the same mass distributions in different
environments.

This observed mass distribution of E galaxies is expected by
hierarchical clustering models, where higher mass objects are
formed in denser regions \citep{Lem99}, and most of them turn out
to be Es in numerical simulations \citep{Spr01}. 
On the other hand, this result can be
interpreted as an evolutionary difference if E are the product of
a faster and older process of growth of spheroids through
dissipationless mergers or coherent collapse, while lenticulars
may be formed at least partially, from stripping of spirals by
strong two-body gravitational interactions \citep{Dre97}. We
should take this fact into account when interpreting the
index$-\log\sigma_0$ relations in the next section.

\subsection{Index$-\log\sigma_0$ Relations}

The first well established relation between an index and
$\sigma_0$ was set for the Mg$_2$ index \citep[e.g.,][]{Ter81,
Dre87, Ben93, Ber98}, which is generally interpreted as a
relationship between mass and metallicity. However, there is some
debate about the true nature of this relation, which might be
influenced by stellar ages or variations in abundances ratios, for
instance that of [$\alpha$/Fe]. In particular, for a given mass,
the small scatter of the Mg$_2-\log\sigma_0$ relation has also been
attributed to a ``conspiracy'' between metallicity and age
\citep{Tra00b}, in the sense that older galaxies are less rich in
metals. Thus, it is important to investigate the
index$-\log\sigma_0$ relations for other elements, whose slopes
and scatter should be sensitive to different chemical composition
variations.

\begin{figure}
   \includegraphics[angle=0,scale=0.6]{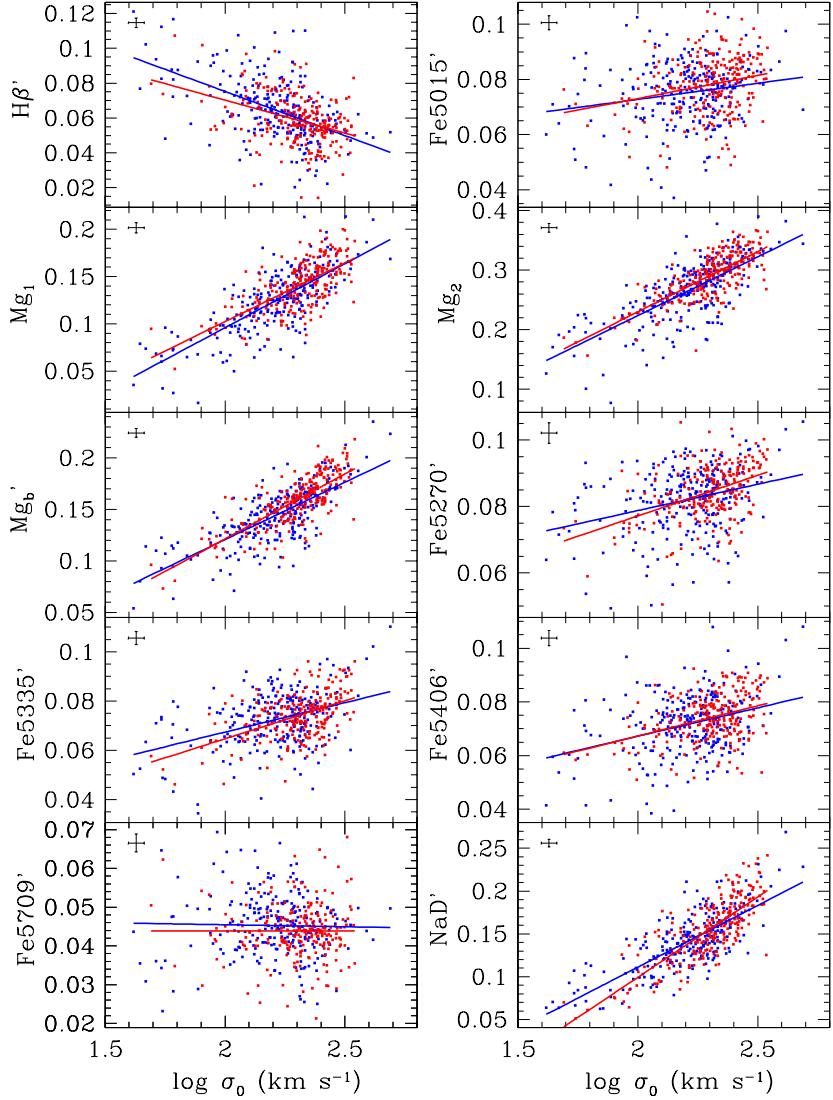}
   \vspace{0.0 cm}
   \caption{Index$-\log\sigma_0$ relations for E (red dots) and S0 (blue dots) galaxies.
The ordinary least-square linear fits to the data are shown as
continuous lines with the same colors as the related dots. Mean
error bars are indicated in the upper left corner of each panel
and indices are expressed in magnitudes. }
   \label{index_sigma_rel}
\end{figure}

In the discussion below, we present all of our measured indices
(I) in magnitudes (I$'$), using the transformation given by
equation (\ref{conv_ind_mag}), in order to permit the comparison
with other authors. The I$'-\log\sigma_0$ relations for 10
measured indices, separating galaxies by morphology (E and S0),
are shown in Figure \ref{index_sigma_rel}. The lines in the Figure
represent the best fit, I$'=A+B\log\sigma_0$, to the data. The
parameters of the fits and their associated errors are presented
in Table \ref{tab:index_sigma_par}, which also gives the number of
galaxies ($N_{gal}$) used in the fit, the Spearman rank $r_s$
coefficient and the $t$ parameter, which tests the null hypothesis
($B=0$). We remind that a value of $t>1.96$ means that the slope ($B$) is significantly different from zero. We stress that our data were
obtained using a long slit and each spectrum represents a
light-weighted measurement, reflecting the mean stellar population
contributing to most of the galaxy's light.

From Figure \ref{index_sigma_rel} and Table
\ref{tab:index_sigma_par} we find that, in general, there is a
significant correlation between the metallic indices and
$\sigma_0$, as indicated by the rank coefficient $r_s$, which is
specially strong for the Mg and NaD lines. The Fe lines, on the
other hand, tend to have a moderate correlation with $\sigma_0$.
The only exception is the Fe5709$'$ index which, in agreement with
\cite{Cle06}, shows no correlation. This behavior is probably due
to the weak dependence of this Lick index on Fe abundance, as
discussed by \citet{Kor05}. We point out that, in contrast to our
result and the one reported by \cite{Nel05}, \cite{Cle06} also finds no
correlation for the Fe5270$'$ index.

The NaD$'$ line shows a correlation with $\sigma_0$ as significant
as that of the Mg lines \citep[see also][]{Cle06}, but its known
dependence on interstellar absorption hampers a reliable
interpretation. 
While S0s may be affected, one should not expect a strong
contribution from the ISM in a ``dust-free'' object like an E
galaxy.

Finally, Figure \ref{index_sigma_rel} shows that the Balmer line
H$\beta '$ is the only one that decreases with increasing
$\sigma_0$ (anti-correlates). This result is consistent with
several previous works \cite[e.g.,][]{Tra98,Den05a,San06a,Cle06}.

\begin{figure}
   \includegraphics[angle=0,scale=0.6]{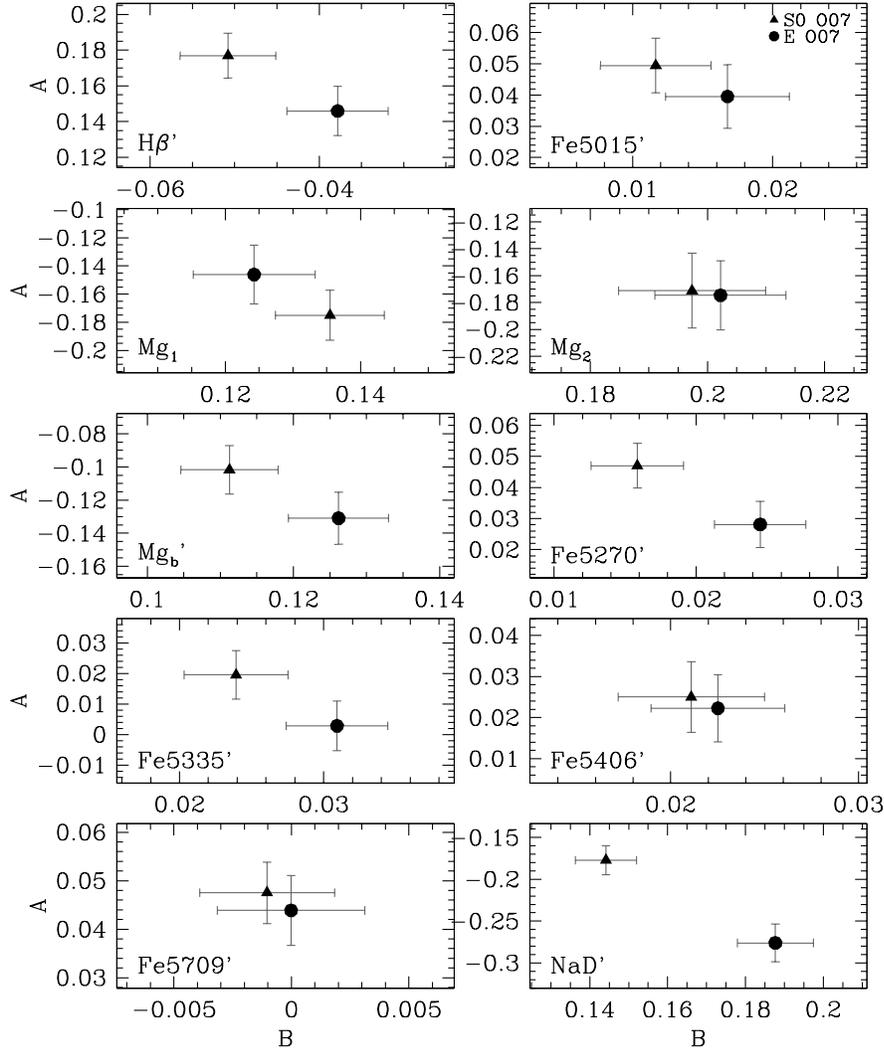}
   \vspace{0.0 cm}
   \caption{Index$-\log\sigma_0$ linear fits (I$'=$A$+$B$\log\sigma_0$) coefficients for
   E (circle) and S0 (triangle) galaxies.}
   \label{index_sigma_par_plot}
\end{figure}

In order to verify possible differences of the I$'-\sigma_0$
relation between E and S0 galaxies, we compare in Figure
\ref{index_sigma_par_plot} the linear fit coefficients (Table
\ref{tab:index_sigma_par}) obtained for these two sub-samples.
From this figure, taking into account the coefficients errors, the
metallic indices do not differ by more than 1.5$\sigma$.
Interestingly, the most discrepant index is the NaD$'$ line, whose
behavior may be caused by the influence of the more conspicuous
interstellar medium expected for the lenticulars. This can also be
seen on the slightly higher H$\beta '$ values for low mass S0,
indicating that this gas reservoir can be used to fuel star
formation.

\begin{figure}
   \includegraphics[angle=0,scale=0.6]{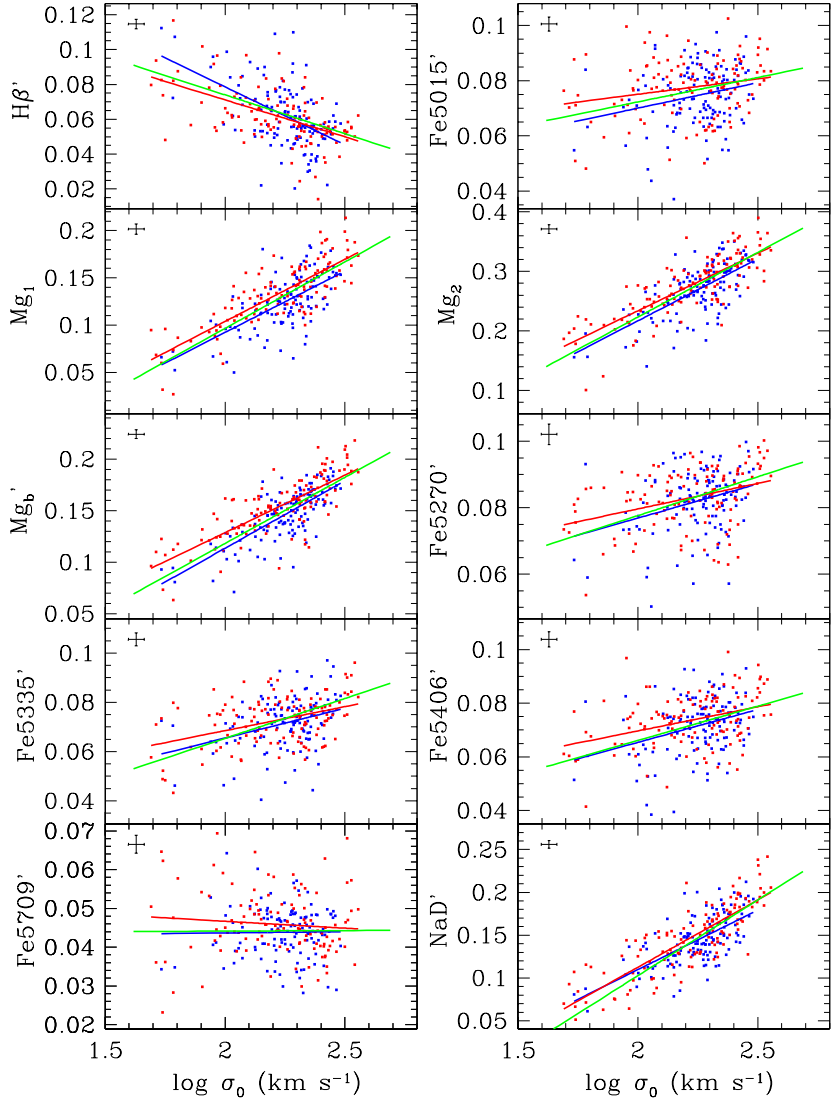}
   \vspace{0.0 cm}
   \caption{Index$-\log\sigma_0$ relations for galaxies in LD (blue dots) and HD (red dots)
   environments. The ordinary least-square linear fits to the data are shown as continuous
   lines with the same colors as the related dots. MD points were omitted for clarity, but
   its fit is shown as a green line. Mean error bars are indicated in the upper left corner
   of each panel and indices are expressed in magnitudes.}
   \label{index_sigma_env}
\end{figure}

Similarly, we examine the possible influence of the environment on
the relations defined above, using the definitions of \S\
\ref{gal_env}, classifying galaxies in regions of low, medium and
high density of objects (LD, MD and HD). Figure
\ref{index_sigma_env} shows the results and the parameters for the
fits are listed in Table \ref{tab:index_sigma_env_par}.
As can be seen, there are no significant differences between metallic
I$'-\log\sigma_0$ relations for LD and HD environments, in agreement with several previous works, except for H$\beta '$. 
Galaxies in LD environments display a steeper relation
than the ones in HD, possibly indicating that the latter have, on
average, older stellar populations. Furthermore, a less steep
relation in high density regions, is an indication of a truncated
star formation history at the low mass end. Possible mechanisms
responsible for this could be harassment \citep{Moo98} or gas
stripping \citep{Gun72}, both of which keep the morphology intact.

\begin{figure}
   \includegraphics[angle=0,scale=0.6]{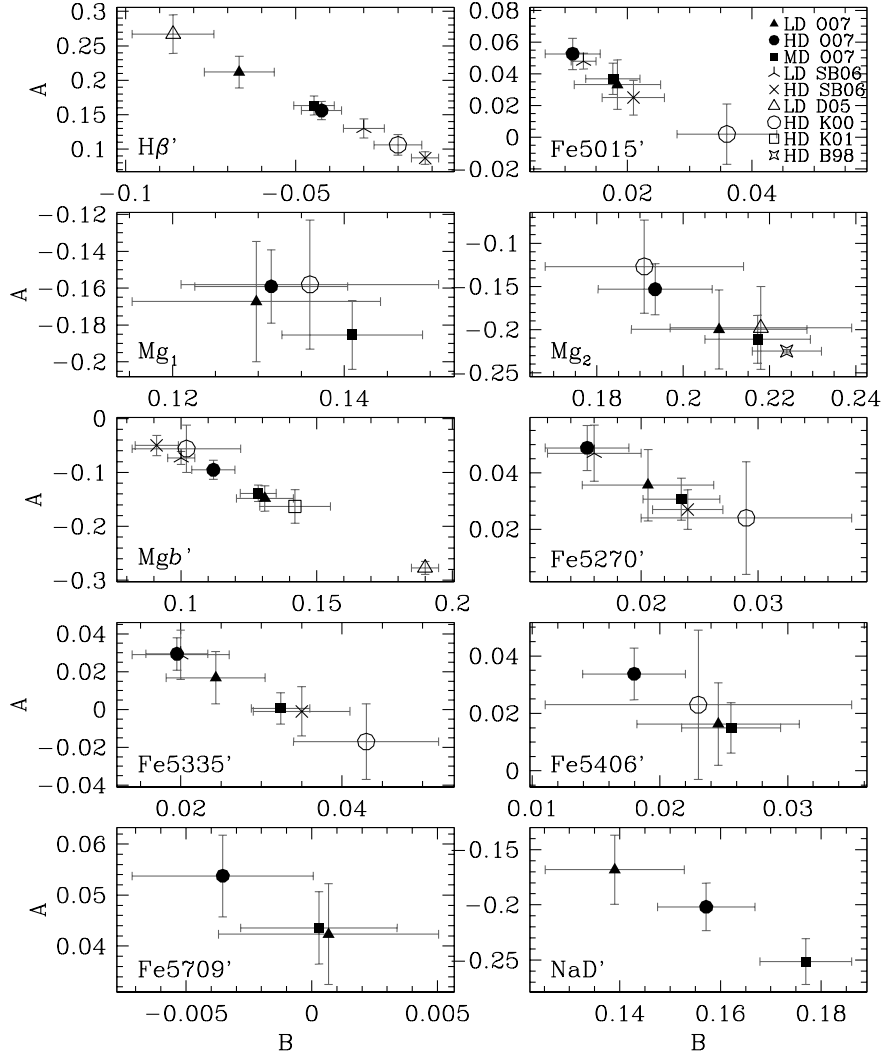}
   \vspace{0.0 cm}
   \caption{Index$-\log\sigma_0$ linear fits (I$'=$A$+$B$\log\sigma_0$) coefficients for
   galaxies in LD (filled triangle), MD (filled square), and HD (filled circle)
   environments. The other symbols displayed are SB06 - \cite{San06a}; D05 - \cite{Den05a};
   K00 - \cite{Kun00}; K01 - \cite{Kun01}; and B98 - \cite{Ber98}. For those samples
   characterized as being made of galaxies of ``isolated'' or ``cluster'' regions, we
   added a LD or HD to the reference code. }
   \label{index_sigma_env_par_plot}
\end{figure}

A comparison of our results with those available in the literature
for nearby samples is shown in Figure
\ref{index_sigma_env_par_plot}, where we plot the coefficients
computed in different environments. This is done despite the
distinct selection criteria of each work,
their reduction and analysis procedures, and density estimator.
In the comparison we include the following authors:
1) \citet{Ber98} who analyzed the Mg$_2$ index for field, group
and cluster galaxies. Their linear fits for each environment are
very similar, thus we show only their cluster fit; 2)
\citet{Kun00} who analyzed data for 11 E and 11 S0 galaxies of the
Fornax cluster, considered here equivalent to our HD definition;
3) \citet{Kun01} who analyzed 72 galaxies in groups and clusters;
4) \citet{Den05a} who analyzed results for 84 galaxies with
indices measured at $r_e/8$, distributed in the field and in
groups, including 8 isolated galaxies; 5) \citet{San06a} who
analyzed I$'-\log\sigma_0$ relations for a sample of 98 objects
divided in two environments defined by low (field, group and Virgo
cluster) and high galaxy density (central region of Coma cluster).

Figure \ref{index_sigma_env_par_plot} shows that within the uncertainties, our I$'-\log\sigma_0$ relations do not depend on environment. This is in
agreement with the findings of \citet{Ber98} and \citet{Den05a}.
In contrast, \citet{San06a} reported a significant difference, and
attributed it to a variation of chemical abundance ratios. These
conflicting results may originate from the $\sigma_0$ ranges
considered by the authors, sometimes by the sparse sampling of low
$\sigma_0$ objects, by poor spectral resolution, or by the
methodology in obtaining and analyzing the data.

We also compare our results with deeper samples. For instance,
\citet{Ber03d} find no dependence of the Mg$_2-\sigma_0$ and
H$\beta '-\sigma_0$ relations on environment using the SDSS data.
By contrast, \citet{Cle06} using data from the same survey, report
``small, but very significant trends with environments''. A
potential problem with analyses based on the SDSS is the adopted
morphological classification, which is based exclusively on color
properties. That may lead to a mixture of populations, including
bulges of early spirals, which may follow different star formation
histories \citep{Tra04}. Since they reside in low-density regions,
this may introduce a false dependence on environment.

The weaker correlation of Fe indices with velocity dispersion, as
compared to those for $\alpha$ representatives (Mg) and the
anti-correlation of H$\beta '$, which is not very sensitive to
$\alpha$/Fe ratio \citep{Kor05}, suggests a variation of the
$\alpha$/Fe ratio with $\sigma_0$. This abundance ratio is
associated to a short star formation history \citep{Tra00,Tho05},
which taken together with their inferred old luminosity-weighted
ages and metal rich content, are conflicting with the predictions
from hierarchical clustering models and give support to the
so-called downsizing scenario \citep{Cow96}, as observed by recent
works \citep[e.g.,][]{Den05a,Tho05,San06a}. Theoretical efforts to
explain these apparent inconsistencies are based on the use of
active galactic nuclei feedback to regulate the star formation
history \citep{Sca05,GdL06}, also separating the stellar
population birth time from the galaxy mass assembly time
\citep{GdL06}. Nevertheless, while no specific abundance ratio
prediction is made, forming the bulk of stars before mass assembly
leads to subsequent ``dry-mergers'', currently a popular scenario
for the origin of bright E galaxies \citep{Ber07}.

Regarding the scatter of the I$'-\log\sigma_0$ relation, although
relatively small, specially in the case of the Mg$_2$ line, it can
not be explained solely by the uncertainties. As we have seen, the
$\alpha$/Fe ratio and age also contribute to the slope of this
relation, and one might wonder if the scatter can also be driven
by age, an usual claim in the literature
\citep[e.g.,][]{Ter02,San06a}, or by variations of abundance
ratios \citep[e.g.,][]{San06a}.

\begin{figure}
   \includegraphics[angle=0,scale=0.6]{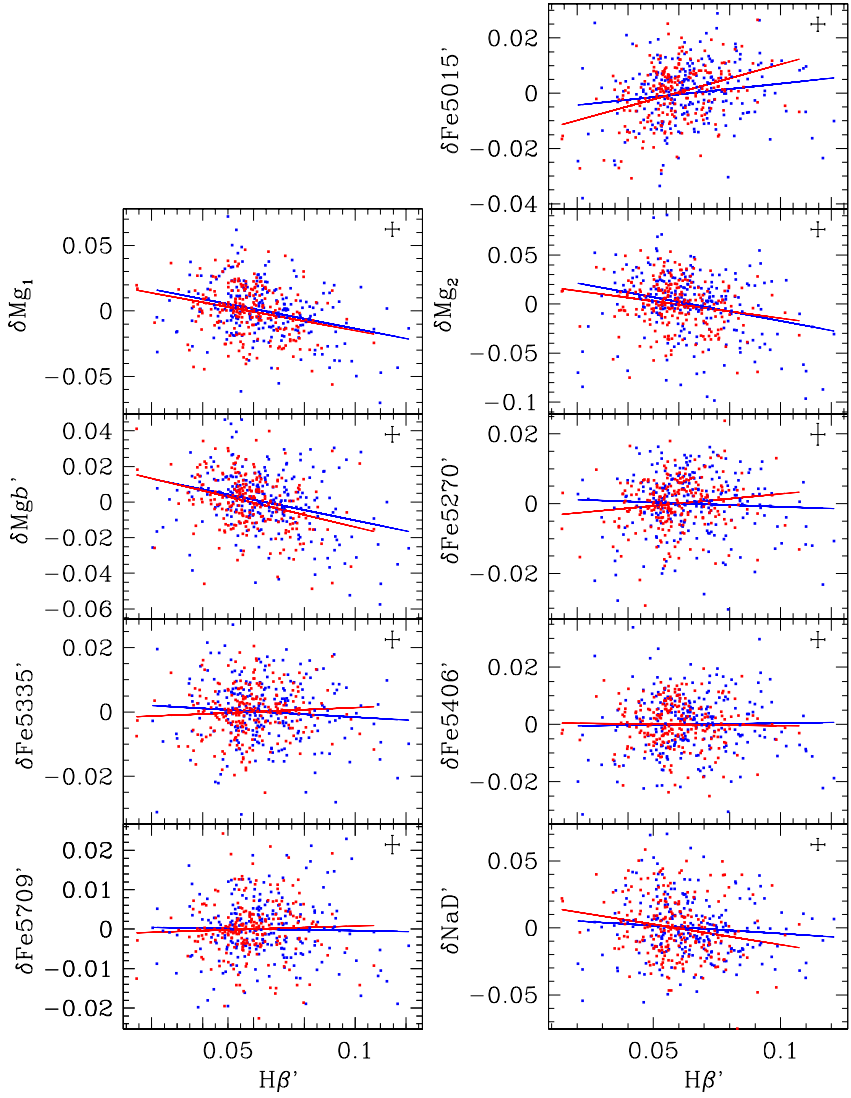}
   \vspace{0.0 cm}
   \caption{Residuals of the index$-\log\sigma_0$ relation versus
   H$\beta '$ for E (red dots) and S0 (blue dots) galaxies.
  In the upper right corner we display the mean error bar. }
   \label{res_index_sigma_morf}
\end{figure}

Following the discussion made by \citet{San06a} we tested for
possible age effects on the scatter of the different
I$'-\log\sigma_0$ relations by measuring the correlation of the
residuals ($\delta$I$'$) of the least-square fits to the relations
versus the H$\beta '$ index, which is an age-sensitive indicator.
The plots of the $\delta$I$'-$H$\beta '$ relations are shown in
the Figure \ref{res_index_sigma_morf} for the sample divided by
morphology, while the correlation parameters are listed in Table
\ref{tab:index_sigma_morf_res} which contains the results of the
Spearman and $t$ tests. Apparently, there is no dependence of the
scatter correlation on morphology (Table
\ref{tab:index_sigma_morf_res}). The lines that show a significant
anti-correlation of the residuals with H$\beta '$ are those of Mg
and NaD$'$. Differently, the Fe (except for Fe5015$'$) residuals
do not show any dependence on H$\beta '$. The fact that the
scatter of distinct metallic lines behave differently with H$\beta
'$ suggests that not only age has an effect on the scatter of the
I$'-\log\sigma_0$ relation, but it may also be affected by the
$[\alpha/$Fe$]$ ratio as suggested by \cite{San06a}. The Fe5015$'$
index, according to the line formation calculations of
\citet{Kor05}, is more sensitive to the chemical abundances of Ti
and Mg than to that of Fe, where in fact, Fe5015$'$
anti-correlates with the abundance of Mg, what could explains its
odd behavior amidst the Fe lines.

\begin{figure}
   \includegraphics[angle=0,scale=0.6]{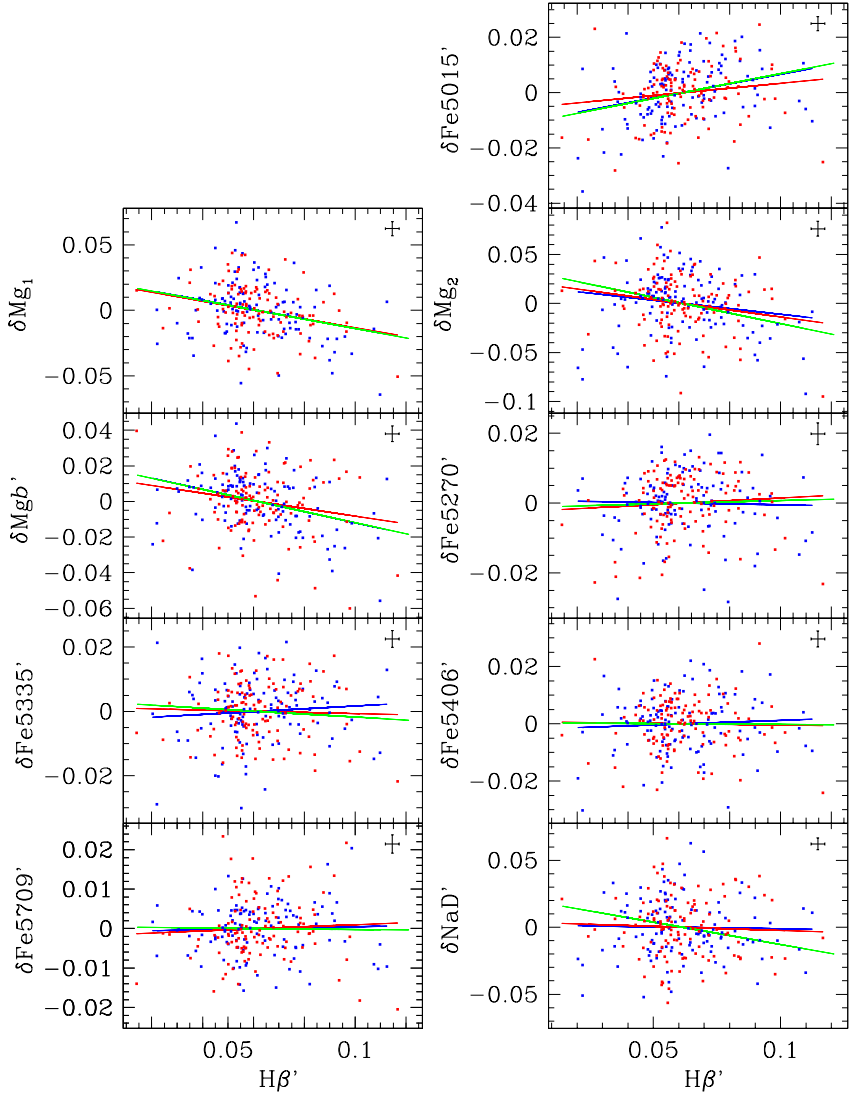}
   \vspace{0.0 cm}
   \caption{Residuals of index$-\log\sigma_0$ relation versus H$\beta '$ for galaxies in
   LD (blue dots) and HD (red dots) environments. Related fits are also shown as continuous
   lines. MD points were omitted for clarity, but its fit is shown as a green
   line. In the upper right corner we display the mean error bar.}
   \label{res_index_sigma_env}
\end{figure}

The results of the $\delta$I$'-$H$\beta '$ relations according to
the environments are shown in Figure \ref{res_index_sigma_env} and
in Table \ref{tab:index_sigma_env_res}. The Mg and Na indices,
that follow the $\alpha$ elements, give distinct answers, but Mg
indices still tend to anti-correlate while the NaD$'$ index shows
no correlation at all, a result that is shared by the Fe indices.
In general, we find that those correlations do not vary
significantly between LD and HD environment, except for the
Fe5015$'$ and the Mg$b$ index. Similar analysis was carried out by
\citet{San06a}, who also found that Mg and Fe lines scatter with
H$\beta '$ behave differently, attributing it to $\alpha$/Fe
variations. However, they find that the residuals of Fe lines with
H$\beta '$ correlate in LD regions, but not in HD. On the other
hand, \citet{Kun02} compared the scatter of the Mg$b$-$\sigma_0$
relation in low density regions and the Fornax cluster and found
no difference between those environments. We also note that the
scatter increases for galaxies with large H$\beta '$, which are
the low mass objects. This result points towards the scenario of
downsizing \citep{Cow96}, where low mass galaxies have an extended
star formation history.


\section{Summary}
\label{disc_conc}

In this work we present the measurements of velocity dispersion
and Lick indices obtained from high $S/N$ ($>20$), long-slit
spectra for 509 early-type galaxies, drawn from the ENEAR survey
\citep{daC00}, considered a fair representation of the present-day
early-type population, having fairly similar distributions of apparent magnitude, radial velocity, morphology and velocity dispersion to the ENEAR sample as a whole. 
This sample is one of the largest of its kind currently available in the nearby Universe ($z<0.002$),  and complements the much larger and higher redshift
($z>0.005$) data from the SDSS \citep{Ber03a,Cle06}. While the
latter has a large number of objects which is unmatched for
statistical studies, it suffers from the uncertainties of
color-based morphological classification, the limitations
inherited to  fiber-based measurements and the need of spectra
stacking, given the typical low $S/N$.

We find that our measurements of velocity dispersion and line
strength indices are, in general, in good agreement with those
available in the literature. A few discrepant cases exist, but
these vary from line to line and from  author to author, showing
no systematic behavior. These data are used in this and
forthcoming papers to study the chemical evolution of early-type
galaxies. In the present paper we use different line
indices-velocity dispersion relations to probe the properties of
early-type galaxies. The relatively large sample allows us to
split it, both by morphological types and different galaxy density
regimes, without compromising the statistical significance of the
sub-samples considered. We emphasize that our densities are
estimated locally, therefore more representative than the broad
categories such as field/groups/clusters normally used in some
studies.

The main findings of this paper are:
\begin{enumerate}

\item Our new measurements of velocity dispersion and Mg$_2$ are
robust, agreeing  with those obtained by  \citet{Weg03} using a
different methodology.

\item Our indices' measurements show small offsets when compared
to those of other authors \citep{Tra98,Tra00,Weg03,Den05a,San06a},
but with comparable scatter, consistent with the error estimates.
Furthermore, several more subtle effects such as the placement of
the slit, the aperture size and the calibration procedure adopted
may affect the measurements.

\item Independent of the mass estimator used we find that E
galaxies in high density regimes are on average more massive than
those located in low-density regions. This is in agreement with
the conclusions of \citet{Cle06}. In addition, we find that E are
more massive than S0 galaxies, a fact that has been used as an
argument in favor of the hierarchical growth of elliptical
galaxies.

\item As well-known, we find that all the Mg indices show a
relatively strong increase with $\sigma_0$ while the Fe indices
depend only mildly on it. These correlations are interpreted by
many authors \citep[e.g.,][]{Ber98} as driven mainly by
metallicity in a coeval population, while others believe that age
also plays an important role in defining this relation and its
scatter \citep{Tra00,San06a}.

\item The observed distinct dependence of Mg and Fe on velocity
dispersion, along with the anti-correlation with velocity
dispersion of the H$\beta '$ index, which has a low sensitivity to
$\alpha$/Fe ratio, suggests that the chemical abundance ratio of
Mg/Fe also varies with mass, as pointed out by several authors
\citep[e.g.,][]{Tra00,Kun02,Tho05,San06a}.

\item In general, the metallic I$'-\sigma_0$ relations show no
dependence on morphology or local density of galaxies. The
exception is the NaD$'$, where the low-mass lenticulars show
stronger line indices than ellipticals with comparable $\sigma_0$.
The H$\beta'-\sigma_0$ also shows a weak dependence on both
morphology and local density. This result contrasts with the
findings of \citet{San06a}, based on a considerably smaller
sample, and claims by \citet{Cle06} using the SDSS data.

\item The H$\beta'$ index decreases with $\sigma_0$ which
suggests, along with the metallic lines, that the last episode of
star formation for massive galaxies took place early in the
galaxies history. Comparing this relation in high and low-density
regions we find that the slope of the relation is significantly
flatter in high-density regions indicating that the star formation
of low mass galaxies has been interrupted by some interaction like
harassment \citep{Moo98} or gas stripping \citep{Gun72}.

\item The residuals of Mg-$\sigma_0$ relation show correlation
with H$\beta'$, decreasing for larger values of  H$\beta'$. This
dependence may indicate that variations in age contribute to the
amplitude of the scatter of the Mg-$\sigma_0$ relation, specially
in the case of low mass objects. No such correlation is seen for
Fe, which may be a hint for Mg/Fe variation taking part in the
scatter.

\end{enumerate}

The fact that massive galaxies have on average high values of Mg,
low values of H$\beta'$, and relatively high Mg/Fe ratio, can be
interpreted as evidence that massive elliptical galaxies are
metal-rich and that the last burst of star formation was brief and
took place in an early phase of their history. The data also shows
that low mass early-type galaxies are younger, metal poorer and
have an extended star formation history, except those in high
density regions, where their star formation has been truncated,
probably due to interactions with the intracluster medium. Taken
together, these evidences favor the currently popular downsizing
model for the formation of early-type galaxies.

It should be pointed out that interpreting the relations between
metallic indices and $\sigma_0$ as a mass-metallicity relation for
a population of coeval objects is not strictly correct and will be
explored in a future paper, where we use evolutionary synthesis
models to take into account the effects of age and abundance
ratios (Ogando et al. 2008a).

Finally, as demonstrated in \citet{Oga05}, an alternative way to
constraint models for the formation of early-type galaxies is to
use the dependence of metallicity with the galactocentric radius,
since the existence of steep gradients indicate a quick formation,
as expected in monolithic-like models. In a forthcoming paper
(Ogando et al. 2008b), we use high-quality data for a sub-sample
of the galaxies presented here to measure these gradients,
extending our previous work.


\acknowledgments

R.L.C.O. acknowledges CNPq Fellowship 141420/2002-2; M.A.G.M. CNPq grants
301366/86-1, 472301/04-7 and 308855/06-0; P.S.P. CNPq grant
301373/86-8.


\newpage
\clearpage

\newpage

\newpage

\newpage

\newpage

\newpage

\newpage

\newpage

\newpage

\newpage

\newpage

\newpage

\newpage

\newpage

\newpage

\newpage

\newpage
\clearpage

\newpage

\newpage

\newpage

\newpage

\newpage

\newpage

\newpage

\newpage
\clearpage


\begin{deluxetable}{lccccr@{$\pm$}lr@{$\pm$}lrr}
  \tablecolumns{11} \tablewidth{0pt} \tabletypesize{\scriptsize}
  \tablecaption{Sample of Galaxies.}
  \tablehead{\ \ \ Name & $\alpha$  & $\delta$ & $T$ & $m_B$ & \multicolumn{2}{c}{$v_r$} &
  \multicolumn{2}{c}{$\log \sigma_0$}  & $\log r_e$ & $N_c$\  \\
     &  (2000)  &   (2000) &   &   &\multicolumn{2}{c}{(km s$^{-1}$)}  &\multicolumn{2}{c}{(km s$^{-1}$)}&
      (kpc)&      \\
\ \ \ \ \ (1) &  (2) & (3) & (4) &  (5) & \multicolumn{2}{c}{(6)}
& \multicolumn{2}{c}{(7)} & (8)\ \ & (9) }
 \startdata
NGC 7832         & 00 06 28.5 & $-$03 42 58 & $-$3 &  13.50 &   6247&12 & 2.340&0.096 &   0.725    &    6 \\
NGC 0050         & 00 14 44.7 & $-$07 20 44 & $-$3 &  12.50 &   5548&15 & 2.413&0.020 &   0.750    &    6 \\
NGC 0113         & 00 26 54.7 & $-$02 30 05 & $-$3 &  14.00 &   4458&19 & 2.185&0.046 &   0.356    &    9 \\
NGC 0125         & 00 28 50.4 & $+$02 50 23 & $-$2 &  13.83 &   5288&18 & 2.064&0.094 &   0.539    &   11 \\
NGC 0155         & 00 34 40.1 & $-$10 46 01 & $-$3 &  14.00 &   6291&21 & 2.250&0.051 &   0.700    &   11 \\
ESO 242G014      & 00 34 32.5 & $-$43 39 22 & $-$2 &  14.00 &   5940&14 & 2.017&0.066 &   0.562    &    1 \\
NGC 0163         & 00 35 59.8 & $-$10 07 19 & $-$3 &  13.50 &   6028&21 & 2.301&0.043 &   0.782    &   16 \\
NGC 0179         & 00 37 46.4 & $-$17 51 01 & $-$5 &  13.90 &   6004&20 & 2.393&0.022 & \nodata    &    1 \\
\enddata
\label{tab:lisgal}
\end{deluxetable}


\begin{deluxetable}{crcccccccc}
\tablecolumns{9} \tablewidth{0pt}
\tabletypesize{\scriptsize}
\tablecaption{Observing Setups} \tablehead{ Setup & N$_{gal}$ &
Detector & Gain & Readout Noise &
Grating & Dispersion & Resolution & Spectral Range\\
       &         &           & e$^-$ ADU$^{-1}$ & e$^-$ &
       l mm$^{-1}$   & \AA\ pixel$^{-1}$ &  \AA          &   \AA \\
  (1) &  (2) & (3) & (4) & (5) & (6) & (7) & (8) & (9) }
  \startdata
1 &  73 & CCD \#24  & 2.9 & 8.0 & 600 & 0.93 & 3.0 & 4500--6300\\
2 &  43 & CCD \#39  & 1.2 & 5.5 & 600 & 1.91 & 6.0 & 3750--7300\\
3 & 348 & CCD \#39  & 1.2 & 5.5 &1200 & 0.98 & 3.0 & 4300--6200\\
4 & 144 & CCD \#38  & 1.6 & 7.1 & 600 & 1.91 & 6.0 & 3750--8000\\
\enddata
\tablecomments{N$_{gal}$ means the number of galaxies with one or
more observations in that configuration. All CCDs have size
2048x2048 except \#38 which is 2688x512. The spatial scale is
0.82$''$/pix, and the slit width 2.5$''$. Setup \#1, used the
second order dispersion of the 600 l mm$^{-1}$ grating.
 }
\label{tab:setups}
\end{deluxetable}


\begin{deluxetable}{lcrr}
  \tablecolumns{4} \tablewidth{0pt}
  \tabletypesize{\scriptsize} \tablecaption{Comparison between our
  $\sigma_0$ measurements with other authors.} \tablehead{Reference
  & N$_{gal}$  &$\langle\delta\sigma_0\rangle$ & \ \ $t$\ \ \ \\
  \ \ (1) &  (2) & (3)\ \ \ & (4) }
  \startdata
 \citet{San06a} &  20  &  $-$3$\pm$17 &0.13 \\
 \citet{Den05a} &  27  & $-$16$\pm$32 &0.89 \\
 \citet{Tra98}  &  75  &  $-$7$\pm$31 &0.60 \\
 \citet{Weg03}  & 432  &  $+$9$\pm$21 &2.11 \\
  \enddata
  \label{tab:vdautcmp}
\end{deluxetable}


\begin{deluxetable}{lcccc}
   \tablecolumns{5} \tablewidth{0pt} \tabletypesize{\scriptsize}
   \tablecaption{Line index definitions} \tablehead{ Index & Bandpass
(\AA) & Blue Pseudocontinua (\AA)
& Red Pseudocontinua (\AA) & Units\\
  \ \ (1) &  (2) & (3) & (4) & (5)}
  \startdata
H$\beta$   &4847.87\quad 4876.62 & 4827.87\quad 4847.87 & 4876.62\quad 4891.62 &\AA \\
Fe5015     &4977.75\quad 5054.00 & 4946.50\quad 4977.75 & 5054.00\quad 5065.25 &\AA \\
Mg$_1$     &5069.12\quad 5134.12 & 4895.12\quad 4957.62 & 5301.12\quad 5366.12 &mag \\
Mg$_2$     &5154.12\quad 5196.62 & 4895.12\quad 4957.62 & 5301.12\quad 5366.12 &mag \\
Mg$b$     &5160.12\quad 5192.62 & 5142.62\quad 5161.37 & 5191.37\quad 5206.37 &\AA \\
Fe5270     &5245.65\quad 5285.65 & 5233.15\quad 5248.15 & 5285.65\quad 5318.15 &\AA \\
Fe5335     &5312.12\quad 5352.12 & 5304.62\quad 5315.87 & 5353.37\quad 5363.37 &\AA \\
Fe5406     &5387.50\quad 5415.00 & 5376.25\quad 5387.50 & 5415.00\quad 5425.00 &\AA \\
Fe5709     &5696.62\quad 5720.37 & 5672.87\quad 5696.62 & 5722.87\quad 5736.62 &\AA \\
NaD        &5876.87\quad 5909.35 & 5860.62\quad 5875.62 & 5922.12\quad 5948.12 &\AA \\
$[$O{\sc{iii}}$]$5007 &4996.85\quad 5016.85 & 4885.00\quad 4935.00 & 5030.00\quad 5070.00 &\AA \\
   \enddata
   \tablecomments{The [O{\sc{iii}}]5007 index is used to estimate the emission level
   in the spectra.}
   \label{tab:line_def}
\end{deluxetable}
%

%
\begin{deluxetable}{lr@{$\pm$}lr@{$\pm$}lr@{$\pm$}lr@{$\pm$}l}
   \tablecolumns{9} \tablewidth{0pt} \tabletypesize{\scriptsize}
   \tablecaption{Averages of Index Gradients}
   \tablehead{  Index &  \multicolumn{2}{c}{$\beta$} \\
 \ \ (1) &  \multicolumn{2}{c}{(2)}}
   \startdata
 H$\beta$ &  0.002 & 0.027 \\
 Fe5015 &  -0.012 & 0.027 \\
 Mg$_1$ &  -0.033 & 0.026\\
 Mg$_2$ &  -0.055 & 0.035\\
 Mg$_b$ &  -0.031 & 0.034\\
 Fe5270 &  -0.016 & 0.025 \\
 Fe5335 &  -0.012 & 0.027 \\
 Fe5406 & -0.015 & 0.029 \\
 Fe5709 & -0.000 & 0.036 \\
 NaD & -0.034 & 0.022 \\
   \enddata
   \label{tab:med_gradind}
\end{deluxetable}
%


\begin{deluxetable}{lrrrrrrrrrrr}
\tablecolumns{1} \tablewidth{0pt} \tabletypesize{\scriptsize}
\tablecaption{Global indices measurements} \tablehead{\ \  Name &
H$\beta$ \   &  Fe5015 & Mg$_1$ & Mg$_2$ & Mg$b$  & Fe5270
& Fe5335  & Fe5406  & Fe5709  & NaD & $[$O {\sc{iii}}$]$  \\
\ \ \ \ (1) & (2)\ \ & (3)\ \ \ & (4)\ \ & (5)\ \ & (6)\ \ & (7)\
\ \ & (8)\ \ \ & (9)\ \ \ & (10)\ \ \ & (11)
 & (12)
} \startdata
NGC 7832         &   1.032    &   5.143    &   0.132    &   0.309    &   4.653    &   2.820    &   2.382    &   1.859    &   0.792    &   4.518    &   0.613    \\
                 &   0.084    &   0.597    &   0.028    &   0.002    &   0.488    &   0.512    &   0.572    &   0.127    &   0.114    &   0.076    &   0.188    \\
NGC 0050         &   1.375    &   5.533    &   0.137    &   0.303    &   4.688    &   2.963    &   2.738    &   1.610    &   0.642    &   4.573    &   0.767    \\
                 &   0.015    &   0.217    &   0.011    &   0.002    &   0.078    &   0.392    &   0.218    &   0.066    &   0.192    &   0.170    &   0.077    \\
NGC 0113         &   1.984    &   4.747    &   0.110    &   0.256    &   4.196    &   2.905    &   2.708    &   1.309    &   1.204    &   3.004    &   0.959    \\
                 &   0.054    &   0.047    &   0.003    &   0.006    &   0.070    &   0.078    &   0.044    &   0.026    &   0.036    &   0.077    &   0.051    \\
NGC 0125         &   1.825    &   4.464    & \nodata    &   0.153    &   3.199    &   2.620    &   2.110    &   1.979    &   0.883    &   3.403    &   0.568    \\
                 &   0.049    &   0.051    & \nodata    &   0.005    &   0.059    &   0.054    &   0.029    &   0.026    &   0.037    &   0.099    &   0.061    \\
\enddata
  \tablecomments{The index [O {\sc{iii}}] in column (12) refers to the
  line of [O {\sc{iii}}]$\lambda$5007. }
  \label{tab:line_indices}
\end{deluxetable}


\begin{deluxetable}{lc@{$\pm$}cccc}
  \tablecolumns{6} \tablewidth{0pt}
  \tabletypesize{\scriptsize} \tablecaption{Fractional error statistics based on
  multiple observations ($N>2$).}
\tablehead{Index & \multicolumn{2}{c}{Average} & Median & Lower quartile &
Upper quartile \\
 \ \ (1) &  \multicolumn{2}{c}{(2)} & (3) & (4) & (5) }
  \startdata
 H$\beta$ &  0.103 &  0.071 &  0.085 &  0.044 & 0.151\\
 Fe5015 &  0.085 &  0.059 &  0.074 &  0.037 & 0.123\\
 Mg$_1$ &  0.062 &  0.051 &  0.048 &  0.023 & 0.093\\
 Mg$_2$ &  0.036 &  0.034 &  0.025 &  0.010 & 0.055\\
 Mg{\it{b}} &  0.046 &  0.031 &  0.045 &  0.020 & 0.063\\
 Fe5270 &  0.071 &  0.053 &  0.053 &  0.038 & 0.096\\
 Fe5335 &  0.081 &  0.054 &  0.071 &  0.042 & 0.117\\
 Fe5406 &  0.087 &  0.055 &  0.076 &  0.052 & 0.115\\
 Fe5709 &  0.112 &  0.071 &  0.103 &  0.048 & 0.162\\
 NaD &  0.045 &  0.035 &  0.037 &  0.022 & 0.058\\
  \enddata
  \tablecomments{The values quoted in the table refer to the distribution
  of final fractional errors distribution. This is done combining several
  measurements for the same galaxy on different occasions or with distinct setups. }
  \label{tab:statmultobs}
\end{deluxetable}


\begin{deluxetable}{lcrc}
\tablecolumns{4} \tablewidth{0pt}
\tabletypesize{\scriptsize}
\tablecaption{Comparison between our galaxies indices measurements and
the literature.}
\tablehead{ Index &  $N_{gal}$ &  $\langle\delta$I$\rangle$\ \ \ \ \ \ &  $t$\ \ \\
 \ \ (1) &  (2) & (3)\ \ \ \ \ \ & (4) }
  \startdata
\multicolumn{4}{l}{\citet{Tra98}}  \\
\tableline
 H$\beta$ &  83 &  -0.030$\pm$0.331 &  0.42 \\
 Fe5015 &  85 &  0.031$\pm$0.728 &  0.24 \\
 Mg$_1$ &  89 &  -0.000$\pm$0.016 &  0.06 \\
 Mg$_2$ &  89 &  0.000$\pm$0.021 &  0.06 \\
 Mg$b$ &  88 &  -0.001$\pm$0.426 &  0.01 \\
 Fe5270 &  90 &  -0.020$\pm$0.389 &  0.33 \\
 Fe5335 &  90 &  0.060$\pm$0.415 &  1.01 \\
 Fe5406 &  83 &  -0.001$\pm$0.377 &  0.03 \\
 Fe5709 &  78 &  0.021$\pm$0.259 &  0.63 \\
 NaD &  90 &  0.248$\pm$0.434 &  1.55 \\
\tableline
\multicolumn{4}{l}{\citet{Tra00}}\\
\tableline
 H$\beta$ &  22 &  0.088$\pm$0.154 & 0.94\\
 Mg$b$ &  24 &  -0.204$\pm$0.239 & 1.20\\
 Fe5270 &  24 &  -0.158$\pm$0.180 & 3.15\\
 Fe5335 &  24 &  -0.028$\pm$0.237 & 0.43\\
\tableline
\multicolumn{4}{l}{\citet{Den05a}}\\
\tableline
 H$\beta$ &  31 &  0.028$\pm$0.282 & 0.16\\
 Fe5015 &  30 &  0.036$\pm$0.709 & 0.19\\
 Mg$_1$ &  30 &  -0.008$\pm$0.017 & 0.99\\
 Mg$_2$ &  30 &  -0.003$\pm$0.018 & 0.29\\
 Mg$b$ &  33 &  -0.314$\pm$0.601 & 1.59\\
 Fe5270 &  32 &  0.113$\pm$0.268 & 1.44\\
 Fe5335 &  33 &  0.241$\pm$0.479 & 2.22\\
 Fe5406 &  32 &  -0.130$\pm$0.254 & 2.16\\
 Fe5709 &  31 &  -0.124$\pm$0.250 & 2.58\\
 NaD &  30 &  -0.012$\pm$0.362 & 0.04\\
\tableline
\multicolumn{4}{l}{\citet{San06a}} \\
\tableline
 H$\beta$ &  20 &  0.131$\pm$0.228 & 1.44\\
 Fe5015 &  22 &  -0.406$\pm$0.536 & 2.59\\
 Mg$b$ &  22 &  -0.411$\pm$0.329 & 1.78\\
 Fe5270 &  22 &  -0.108$\pm$0.231 & 1.44\\
 Fe5335 &  22 &  -0.043$\pm$0.281 & 0.49\\
\tableline \multicolumn{4}{l} {\citet{Weg03}}\\
 \tableline
Mg$_2$ & 418 & -0.001$\pm$0.024 & 0.42 \\
\enddata
\label{tab:comp_todos_autores}
\end{deluxetable}


\begin{deluxetable}{lrcccccc}
   \tablecolumns{8} \tablewidth{0pt} \tabletypesize{\scriptsize}
   \tablecaption{Mean values of mass indicators.}
   \tablehead{ Sample & $N_{gal}$ & $\langle\log \sigma_0\rangle$ & Skewness &
  Kurtosis & $\langle\log M_e\rangle$ & Skewness & Kurtosis \\
 &         &  (km\ s$^{-1}$) &  &  & (M$_{\odot}$) & & \\
 (1) &  (2) & (3) & (4) & (5) &  (6) & (7) & (8) }
    \startdata
 E Total  &  226\ \ &  2.30$\pm$0.15 & -0.38 &  1.92 & 10.79$\pm$0.52 & -0.32 &  0.86 \\
 S0 Total &  230\ \ &  2.21$\pm$0.18 & -0.43 &  1.02 & 10.51$\pm$0.59 & -0.06 &  0.01 \\
 E LD     &   49\ \ &  2.28$\pm$0.13 & -0.38 &  3.24 & 10.73$\pm$0.49 & -0.45 &  1.97 \\
 E MD     &  116\ \ &  2.32$\pm$0.12 & -0.12 & -0.07 & 10.85$\pm$0.43 &  0.14 & -0.11 \\
 E HD     &   61\ \ &  2.27$\pm$0.20 & -0.87 &  0.20 & 10.72$\pm$0.67 & -0.77 & -0.45 \\
 S0 LD    &   71\ \ &  2.22$\pm$0.15 & -0.09 &  1.12 & 10.52$\pm$0.52 & -0.02 & -0.55 \\
 S0 MD    &   98\ \ &  2.21$\pm$0.18 & -0.72 &  1.80 & 10.53$\pm$0.58 & -0.19 &  0.53 \\
 S0 HD    &   61\ \ &  2.18$\pm$0.20 & -0.44 & -0.35 & 10.46$\pm$0.69 & -0.06 & -0.52 \\
  \enddata
  \label{tab:mass_gal}
\end{deluxetable}


\begin{deluxetable}{lccccc}
    \tablecolumns{6} \tablewidth{0pt} \tabletypesize{\scriptsize}
    \tablecaption{Probabilities of Kolmogorov-Smirnov test between distributions of
    mass indicators.}
    \tablehead{ Sample & $\log\sigma_0$ (\%) & $\log M_e$ (\%) }
    \startdata
S0 - E      & 0 &  0 \\
E: LD - HD  & 8 &  9 \\
S0: LD - HD & 9 & 40 \\
    \enddata
    \label{tab:cmplvd_gal}
\end{deluxetable}


\begin{deluxetable}{lr@{$\pm$}lr@{$\pm$}lrrr}
  \tablecolumns{8} \tablewidth{0pt} \tabletypesize{\scriptsize}
  \tablecaption{Coefficients of the I$'-\log \sigma_0$ linear fits for
  different morphological types.}
  \tablehead{ Index & \multicolumn{2}{c}{A} &
  \multicolumn{2}{c}{B} & $N_{gal}$ & $r_s$ \ \ & $t$\ \ }
  \startdata
& \multicolumn{7}{c}{Elliptical galaxies}\ \ \ \ \\
\hline
 H$\beta$ &  0.146&0.014 &  -0.038&0.006 &  231 &  -0.36 & -5.82\\
 Fe5015 &  0.040&0.010 &  0.017&0.004 &  234 &  0.23 & 3.68\\
 Mg$_1$ &  -0.146&0.021 &  0.124&0.009 &  231 &  0.65 & 12.92\\
 Mg$_2$ &  -0.175&0.026 &  0.202&0.011 &  237 &  0.72 & 16.11\\
 Mg$b$ &  -0.131&0.016 &  0.126&0.007 &  238 &  0.74 & 17.11\\
 Fe5270 &  0.028&0.007 &  0.025&0.003 &  237 &  0.42 & 7.11\\
 Fe5335 &  0.003&0.008 &  0.031&0.004 &  236 &  0.44 & 7.42\\
 Fe5406 &  0.022&0.008 &  0.023&0.004 &  235 &  0.32 & 5.21\\
 Fe5709 &  0.044&0.007 &  -0.000&0.003 &  230 &  0.00 & 0.03\\
 NaD &  -0.276&0.023 &  0.188&0.010 &  238 &  0.76 & 17.98\\
\hline
& \multicolumn{7}{c}{Lenticular galaxies}\ \ \ \ \\
\hline
 H$\beta$ &  0.177&0.013  &  -0.051&0.006 &  238 &  -0.51 & -9.04\\
 Fe5015   &  0.049&0.009  &  0.012&0.004  &  246 &  0.13 & 2.03\\
 Mg$_1$   &  -0.175&0.018 &  0.135&0.008  &  230 &  0.71 & 15.25\\
 Mg$_2$   &  -0.171&0.028 &  0.197&0.013  &  247 &  0.68 & 14.49\\
 Mg$b$   &  -0.102&0.015 &  0.111&0.007  &  245 &  0.69 & 14.80\\
 Fe5270   &  0.047&0.007  &  0.016&0.003  &  248 &  0.23 & 3.76\\
 Fe5335   &  0.020&0.008  &  0.024&0.004  &  247 &  0.27 & 4.41\\
 Fe5406   &  0.025&0.009  &  0.021&0.004  &  246 &  0.21 & 3.34\\
 Fe5709   &  0.048&0.006  &  -0.001&0.003 &  239 &  -0.08 & -1.30\\
 NaD      &  -0.177&0.017 &  0.144&0.008  &  247 &  0.72 & 16.09\\
  \enddata
  \label{tab:index_sigma_par}
\end{deluxetable}


\begin{deluxetable}{lr@{$\pm$}lr@{$\pm$}lrrr}
  \tablecolumns{8} \tablewidth{0pt} \tabletypesize{\scriptsize}
  \tablecaption{Results from the I$'-\log \sigma_0$ linear fits for
  different environments.} \tablehead{ Index &
  \multicolumn{2}{c}{A} & \multicolumn{2}{c}{B} & $N_{gal}$ & $r_s$\ \ &
  $t$\ \ \ }
  \startdata
& \multicolumn{7}{c}{Low Density}\ \ \ \\
\hline
 H$\beta$ &  0.212&0.023 &  -0.067&0.010 & 123 & -0.45 & -5.53\\
 Fe5015 &  0.033&0.016   &  0.018&0.007 &  124 &  0.17 & 1.87\\
 Mg$_1$ &  -0.167&0.033  &  0.130&0.014 &  117 &  0.61 & 8.23\\
 Mg$_2$ &  -0.200&0.046  &  0.208&0.020 &  125 &  0.63 & 9.05\\
 Mg$b$ &  -0.148&0.024  &  0.131&0.010 &  124 &  0.72 & 11.31\\
 Fe5270 &  0.036&0.013   &  0.021&0.006 &  125 &  0.23 & 2.62\\
 Fe5335 &  0.017&0.014   &  0.024&0.006 &  123 &  0.30 & 3.51\\
\enddata
  \label{tab:index_sigma_env_par}
\end{deluxetable}


\begin{deluxetable}{lrrrr}
  \tablecolumns{5} \tablewidth{0pt} \tabletypesize{\scriptsize}
  \tablecaption{Correlation test for the I$'-\log \sigma_0$ residuals versus H$\beta '$ for different morphologies.} \tablehead{ & \multicolumn{2}{c}{S0} & \multicolumn{2}{c}{E} \\
\hline
Index & \multicolumn{1}{c}{$r_s$} & \multicolumn{1}{c}{$t$} & \multicolumn{1}{c}{$r_s$} & \multicolumn{1}{c}{$t$} }
\startdata
 Fe5015 &  0.166 &  2.566 &  0.322 & 5.105\\
 Mg$_1$ &  -0.252 &  -3.843 &  -0.241 & -3.723\\
 Mg$_2$ &  -0.258 &  -4.089 &  -0.183 & -2.812\\
 Mg$b$ &  -0.258 &  -4.082 &  -0.285 & -4.505\\
 Fe5270 &  -0.019 &  -0.296 &  0.157 & 2.395\\
 Fe5335 &  -0.070 &  -1.068 &  0.087 & 1.321\\
 Fe5406 &  0.044 &  0.672 &  -0.006 & -0.089\\
 Fe5709 &  -0.007 &  -0.107 &  0.099 & 1.487\\
 NaD &  -0.137 &  -2.117 &  -0.189 & -2.915\\
\enddata
  \label{tab:index_sigma_morf_res}
\end{deluxetable}


\begin{deluxetable}{lrrrrrr}
  \tablecolumns{7} \tablewidth{0pt} \tabletypesize{\scriptsize}
  \tablecaption{Correlation test for the I$'-\log \sigma_0$ residuals versus H$\beta '$ for
  different environments.} \tablehead{ & \multicolumn{2}{c}{LD} & \multicolumn{2}{c}{MD} & \multicolumn{2}{c}{HD} \\
\hline
Index & \multicolumn{1}{c}{$r_s$} & \multicolumn{1}{c}{$t$} & \multicolumn{1}{c}{$r_s$} & \multicolumn{1}{c}{$t$} & \multicolumn{1}{c}{$r_s$} & \multicolumn{1}{c}{$t$}}
\startdata
 Fe5015 &  0.311 &  3.571 &  0.230 &  3.507 &  0.144 & 1.575\\
 Mg$_1$ &  -0.251 &  -2.744 &  -0.274 &  -4.185 &  -0.242 & -2.631\\
 Mg$_2$ &  -0.175 &  -1.951 &  -0.307 &  -4.818 &  -0.135 & -1.469\\
 Mg$b$ &  -0.323 &  -3.725 &  -0.301 &  -4.723 &  -0.115 & -1.263\\
 Fe5270 &  0.044 &  0.480 &  0.051 &  0.756 &  0.118 & 1.288\\
 Fe5335 &  0.098 &  1.073 &  -0.081 &  -1.209 &  0.026 & 0.282\\
 Fe5406 &  0.015 &  0.163 &  0.022 &  0.327 &  0.032 & 0.343\\
 Fe5709 &  0.070 &  0.744 &  -0.007 &  -0.098 &  0.145 & 1.566\\
 NaD &  -0.045 &  -0.495 &  -0.238 &  -3.649 &  -0.044 & -0.476\\
\enddata
  \label{tab:index_sigma_env_res}
\end{deluxetable}



\begin{thebibliography}{100}

\bibitem[Alonso et al.(2003)]{Alo03} Alonso, M.V., Bernardi, M.,
da Costa, L.N., Wegner, G., Willmer, C.N.A., Pellegrini, P.S., \&
Maia, M.A.G. 2003, \aj, 125, 2307

\bibitem[Bender, Burstein \& Faber(1993)]{Ben93} Bender, R., Burstein, D.,
\& Faber, S.M. 1993, \apj, 411, 153

\bibitem[Bernardi et al.(1998)]{Ber98} Bernardi, M., Renzini, A., da Costa,
L.N., Wegner, G., Alonso, M.V., Pellegrini, P.S., Rit\'e, C., \& Willmer,
C.N.A. 1998, \apj, 508, L143

\bibitem[Bernardi et al.(2002)]{Ber02} Bernardi, M., Alonso, M.V., da Costa,
L.N., Willmer, C.N.A. Wegner, G., Pellegrini, P.S., Rit\'e, C. \& Maia, M.A.G.
2002, \aj, 123, 2159

\bibitem[{{Bernardi} {et~al.}(2003){Bernardi}, {Sheth}, {Annis}, {et~al.}}]{Ber03a}
{Bernardi}, M., {Sheth}, R.K., {Annis}, J., {et~al.} 2003, \aj,
125, 1817

\bibitem[{{Bernardi} {et~al.}(2003){Bernardi}, {Sheth}, {Annis}, {et~al.}}]{Ber03d}
{Bernardi}, M., {Sheth}, R.K., {Annis}, J., {et~al.} 2003, \aj,
125, 1882

\bibitem[Bernardi et al.(2005)]{Ber05} Bernardi, M. Sheth, R.K., Nichol, R.C.,
Schneider, D.P. \& Brinkmann, J. 2005 \aj, 129, 61

\bibitem[{{Bernardi} {et~al.}(2007){Bernardi}, {Hyde}, {Sheth}, {Miller}, \&
{Nichol}}]{Ber07} {Bernardi}, M., {Hyde}, J.B., {Sheth}, R.K.,
{Miller}, C.J., \& {Nichol}, R.C. 2007, \aj, 133, 1741

\bibitem[Burstein et al.(1984)]{Bur84} Burstein, D., Faber, S.M., Gaskell, C.M.
\& Krumm, N. 1984, \apj, 287, 586

\bibitem[Burstein et al.(1997)]{Bur97} Burstein, D., Bender, R., Faber, S.M.,
\& Nolthenius, R. 1997, \aj, 114, 1365

\bibitem[Clemens et al.(2006)] {Cle06} Clemens, M.S., Bressan, A., Nikolic, B.,
Alexander, P., Annibali, F. \& Rampazzo, R. 2006, \mnras, 370, 702

\bibitem[{{Colless} {et~al.}(2001){Colless}, {Dalton}, {Maddox}, {et~al.}}]{Col01}
{Colless}, M., {Dalton}, G., {Maddox}, S., {et~al.} 2001, \mnras, 328, 1039

\bibitem[{{Cowie} {et~al.}(1996){Cowie}, {Songaila}, {Hu}, \&  {Cohen}}]{Cow96}
{Cowie}, L.L., {Songaila}, A., {Hu}, E.M., \& {Cohen}, J.G. 1996,
\aj, 112,  839

\bibitem[{{da Costa} {et~al.}(1998){da Costa}, {Willmer}, {Pellegrini},  {et~al.}}]{daC98}
{da Costa}, L.N., {Willmer}, C.N.A., {Pellegrini}, P.S., {et~al.}
1998, \aj, 116, 1

\bibitem[da Costa et al.(2000)] {daC00} da Costa, L.N., Bernardi, M., Alonso, M.V.,
Wegner, G., Willmer, C.N.A., Pellegrini, P.S., Rit\'e, C., \& Maia, M.A.G. 2000,
\aj, 120, 95

\bibitem[Davies et al.(1987)]{Dav87} Davies, R.L., Burstein, D., Dressler, A.,
Faber, S.M., Lynden-Bell, D., Terlevich, R.J., \& Wegner, G. 1987, \apjs, 64, 581

\bibitem[{{De Lucia} {et~al.}(2006){De Lucia}, {Springel}, {White},  {et~al.}}]{GdL06}
{De Lucia}, G., {Springel}, V., {White}, S.D.M., {et~al.} 2006,
\mnras, 366, 499

\bibitem[Denicol\'o et al.(2005)]{Den05a} Denicol\'o, G., Terlevich, R., Terlevich,
E., Forbes, D.A., Terlevich, A., \& Carrasco, L. 2005, \mnras, 356, 1440

\bibitem[Djorgovski \& Davis (1987)]{DjDv87} Djorgovski, S. \& Davis, M. 1987, \aj, 313, 59

\bibitem[Dressler(1980)]{Dre80} Dressler, A. 1980, \apj, 236, 351

\bibitem[Dressler et al.(1987)]{Dre87} Dressler, A., Lynden-Bell, D., Burstein, D.,
Davies, R.L., Faber, S.M., Terlevich, R., \& Wegner, G. 1987, \apj, 313, 42

\bibitem[Dressler et al.(1997)]{Dre97} Dressler, A., Oemler, A., Warrick, J.C.,
Smail, I., Ellis, R.S., Barger, A., Butcher, H., Poggianti, B.M. \&\ Sharples,
R.M. 1997, \apj, 490, 577

\bibitem[Faber \& Jackson(1976)]{Fab76} Faber, S.M., \& Jackson, R.E. 1976, \apj, 204, 668

\bibitem[Faber et al.(1985)]{Fab85} Faber, S.M., Friel, E.D., Burstein, D.,
\& Gaskell, C.M. 1985, \apjs, 57, 711

\bibitem[Gonz\'alez(1993)]{Gon93} Gonz\'alez, J.J. 1993, PhD Thesis, University
of California, Santa Cruz, USA

\bibitem[Gorgas et al.(1990)]{Gor90} Gorgas, J., Efstathiou, G., \& Arag\'on-Salamanca,
A. 1990, \mnras, 245, 217

\bibitem[{{Gunn} \& {Gott}(1972)}]{Gun72} {Gunn}, J.E. \& {Gott}, J.R.I. 1972, \apj, 176, 1

\bibitem[J\o rgensen, Franx \& Kj\ae rgaard(1995)]{Jor95} J\o rgensen, I.,
Franx, M., Kj\ae rgaard, P. 1995, \mnras, 276, 1341

\bibitem[{{Kauffmann} {et~al.}(1993)}]{Kau93}{Kauffmann}, G., {White}, S.D.M., \&
{Guiderdoni}, B. 1993, \mnras, 264, 201

\bibitem[{{Korn} {et~al.}(2005){Korn}, {Maraston}, \& {Thomas}}]{Kor05} {Korn}, A.J.,
{Maraston}, C., \& {Thomas}, D. 2005, \aap, 438, 685

\bibitem[Kuntschner(2000)]{Kun00} Kuntschner, H. 2000, \mnras, 315, 184

\bibitem[Kuntschner et al.(2001)]{Kun01} Kuntschner, H., Lucey, J.R., Smith, R.J.,
Hudson, M.J., \& Davies, R.L. 2001, \mnras, 323, 615

\bibitem[{{Kuntschner} {et~al.}(2002){Kuntschner}, {Smith}, {Colless},  {et~al.}}]{Kun02}
{Kuntschner}, H., {Smith}, R.J., {Colless}, M., {et~al.} 2002,
\mnras, 337, 172

\bibitem[Kurtz \& Mink(1998)]{Kur98} Kurtz, M.J., \& Mink, D.J. 1998, \pasp, 110, 934

\bibitem[Larson(1974)]{Lar74} Larson, R.B. 1974, \mnras, 166, 585

\bibitem[Lasker et al.(1990)]{Las90} Lasker, B.M., Sturch, C.R., McLean, B.J.,
Russell, J.L., Jenkner, H., Shara, M.M. 1990, \aj, 99, 2019

\bibitem[Lemson \& Kauffmann(1999)]{Lem99} Lemson, G., \& Kauffmann, G. 1999,
\mnras, 302, 111

\bibitem[{{Macchetto} {et~al.}(1996){Macchetto}, {Pastoriza}, {Caon},  {et~al.}}]{Mac96}
{Macchetto}, F., {Pastoriza}, M., {Caon}, N., {et~al.} 1996, \aaps, 120, 463

\bibitem[Maia \& da Costa(1990)]{Mai90} Maia, M.A.G., \& da Costa, L.A.N. 1990,
\apj, 352, 457

\bibitem[{{Moore} {et~al.}(1998){Moore}, {Lake}, \& {Katz}}]{Moo98} {Moore}, B.,
{Lake}, G., \& {Katz}, N. 1998, \apj, 495, 139

\bibitem[Mulchaey \& Zabludoff(1999)]{Mul99} Mulchaey, J.S., \& Zabludoff, A.I. 1999,
\apj, 514, 133

\bibitem[Nelan et al.(2005)]{Nel05} Nelan, J.E., Smith, R.J., Hudson, M.J., Wegner,
G.A., Lucey, J.R., Moore, S.A.W., Quinney, S.J. \& Suntzeff, N.B., \apj,632, 137

\bibitem[Ogando et al.(2005)]{Oga05} Ogando, R.L.C., Maia, M.A.G., Chiappini, C.,
Pellegrini, P.S., Schiavon, R., \& da Costa, L.N. 2005, \apj, 632, L61

\bibitem[Ogando et al.(2008a)]{Oga08a} Ogando, R.L.C., Maia, M.A.G., Pellegrini, P.S.,
\& da Costa, L.N. 2008, \apj, in press

\bibitem[{{Paturel} {et~al.}(2003){Paturel}, {Petit}, {Prugniel}, {et~al.}}]{Pat03}
{Paturel}, G., {Petit}, C., {Prugniel}, P., {et~al.} 2003, \aap, 412, 45

\bibitem[{{Pipino} \& {Matteucci}(2004)}]{Pip04}{Pipino}, A. \& {Matteucci}, F. 2004,
\mnras, 347, 968

\bibitem[Reda et al.(2004)]{Red04} Reda, F.M., Forbes, D.A., Beasley, M.A., O'Sullivan,
E.J., \& Goudfrooij, P. 2004, \mnras, 354, 851

\bibitem[S\'anchez-Bl\'azquez et al.(2006)]{San06a} S\'anchez-Bl\'azquez, P., Gorgas, J.,
Cardiel, N., \& Gonz\'alez, J.J. 2006, \aap, 457, 787

\bibitem[{{Scannapieco} {et~al.}(2005){Scannapieco}, {Silk}, \& {Bouwens}}]{Sca05}
{Scannapieco}, E., {Silk}, J., \& {Bouwens}, R. 2005, \apjl, 635, L13

\bibitem[{{Schiavon} {et~al.}(2002){Schiavon}, {Faber}, {Castilho}, {et~al.}}]{Schi02i}
{Schiavon}, R.P., {Faber}, S.M., {Castilho}, B.V., {et~al.} 2002,
\apj, 580, 850

\bibitem[Springel et al.(2001)]{Spr01} Springel, V., White, S.D.M., Tormen, G.,
\& Kauffmann, G. 2001, \mnras, 328, 726

\bibitem[Terlevich et al(1981)]{Ter81} Terlevich, R., Davies, R.L., Faber, S.M.,
Burstein, D. 1981, \mnras, 196, 381

\bibitem[{{Terlevich} \& {Forbes}(2002)}]{Ter02} {Terlevich}, A.I. \& {Forbes}, D.A.
2002, \mnras, 330, 547

\bibitem[{{Thomas} {et~al.}(2003){Thomas}, {Maraston}, \& {Bender}}]{Tho03}
{Thomas}, D., {Maraston}, C., \& {Bender}, R. 2003, \mnras, 339, 897

\bibitem[Thomas et al.(2005)]{Tho05} Thomas, D., Maraston, C., Bender, R. \&
Oliveira, C.M. \apj, 621, 673

\bibitem[Tonry \& Davis(1979)]{Ton79} Tonry, J., \& Davis, M. 1979, \aj, 84, 1511

\bibitem[Trager et al.(1998)]{Tra98} Trager, S.C., Worthey, G., Faber, S.M.,
Burstein, D., Gonz\'alez, J.J. 1998, \apjs, 116, 1

\bibitem[Trager et al.(2000a)]{Tra00} Trager, S.C., Faber, S.M., Worthey, G.,
Gonz\'alez, J.J. 2000a, \aj, 119, 1645

\bibitem[Trager et al.(2000b)]{Tra00b} Trager, S.C., Faber, S.M., Worthey, G.,
Gonz\'alez, J.J. 2000b, \aj, 120, 165

\bibitem[{{Trager}(2004)}]{Tra04}{Trager}, S.C. 2004, in Origin and Evolution of
the Elements, ed.  A. {McWilliam} \& M. {Rauch}, 388

\bibitem[Wegner et al.(2003)]{Weg03} Wegner, G., Bernardi, M., Willmer, C.N.A.,
da Costa, L.N., Alonso, M.V., Pellegrini, P., Maia, M.A.G., Chaves, O.L.,
\& Rit\'e, C. 2003, \aj, 126, 2268

\bibitem[{{Worthey} {et~al.}(1992){Worthey}, {Faber}, \& {Gonzalez}}]{Wor92}{Worthey},
G., {Faber}, S.M., \& {Gonzalez}, J.J. 1992, \apj, 398, 69

\bibitem[{{Worthey}(1994)}]{Wor94}{Worthey}, G. 1994, \apjs, 95, 107

\bibitem[Worthey \& Ottaviani(1997)]{Wor97} Worthey, G., \& Ottaviani, D.L. 1997,
\apjs, 111, 377

\end{thebibliography}
\end{document}